\documentclass[12pt]{article}
\usepackage{amscd,amssymb,amsmath,latexsym,enumerate}
\usepackage[mathscr]{euscript}
\usepackage{epsfig}
\usepackage{fancybox}
\usepackage{verbatim}
\usepackage{multirow}
\usepackage{stmaryrd}

\usepackage[hidelinks]{hyperref}
\hypersetup{ 
       colorlinks=true,
       linkcolor=black, 
       citecolor=black,
       filecolor=black,
       menucolor=black,
       urlcolor=black,
       breaklinks=true
}

\usepackage{color}

\title{Skew localizer and $\ZM_2$-flows for real index pairings}

\author{Nora Doll, Hermann Schulz-Baldes
\\
\\
{\small Department Mathematik, Friedrich-Alexander-Universit\"at Erlangen-N\"urnberg, Germany}
}

\date{ }

\newtheorem{theo}{Theorem}
\newtheorem{defini}[theo]{Definition}
\newtheorem{proposi}[theo]{Proposition}
\newtheorem{lemma}[theo]{Lemma}
\newtheorem{coro}[theo]{Corollary}

\newcommand{\CM}{{\mathbb C}}

\newcommand{\RM}{{\mathbb R}}

\newcommand{\ZM}{{\mathbb Z}}
\newcommand{\PM}{{\mathbb P}}

\newcommand{\Bb}{{\cal B}}

\newcommand{\Ss}{{\cal S}}

\newcommand{\Rr}{{\cal R}}

\newcommand{\Cc}{{\cal C}}

\newcommand{\Hh}{{\cal H}}

\newcommand{\one}{{\bf 1}}

\newcommand{\SF}{{\rm Sf}}

\newcommand{\Ind}{{\rm Ind}} 
\newcommand{\Ker}{{\rm Ker}} 
 
\newcommand{\Ran}{{\rm Ran}} 
\newcommand{\sgn}{{\rm sgn}} 
\newcommand{\Sig}{{\rm Sig}} 
\newcommand{\diag}{{\rm diag}}

\newcommand{\HSF}{{\mbox{\rm HSf}_2}}
\newcommand{\thetaHSFeven}{\theta_{\rm ev}}
\newcommand{\thetaHSFodd}{\theta_{\rm od}}
\newcommand{\thetaSFeven}{\Theta}

\newcommand{\Pf}{{\rm Pf}}
\newcommand{\Of}{{\rm Of}}
\newcommand{\odd}{{\rm od}}
\newcommand{\even}{{\rm ev}}

\newcommand{\SymProd}{{\mathcal S}}

\newcommand{\Red}{} 

\begin{document}

\maketitle

\begin{abstract} 
Real index pairings of projections and unitaries on a separable Hilbert space with a real structure are defined when the projections and unitaries fulfill symmetry relations invoking the real structure, namely projections can be real, quaternionic, even or odd Lagrangian and unitaries can be real, quaternionic, symmetric or anti-symmetric. There are $64$ such real index pairings of real $K$-theory with real $K$-homology. For $16$ of them, the index of the \Red{Fredholm operator representing the} pairing vanishes, but there is a secondary $\ZM_2$-valued invariant. The first set of results provides index formulas expressing each of these $16$ $\ZM_2$-valued pairings as either an orientation flow or a half-spectral flow. The second and main set of results constructs the skew localizer for a pairing stemming from \Red{an unbounded} Fredholm module and shows that the $\ZM_2$-invariant can be computed as the sign of \Red{the Pfaffian of the skew localizer} and in $8$ of the cases as the sign of the determinant of \Red{the off-diagonal entry of the skew localizer}. This is of relevance for the numerical computation of invariants of topological insulators. 
\end{abstract}


\section{Introduction}
\label{sec-intro}

Index theorems for $\ZM_2$-invariants in differential topology such as the Stiefel-Whitney class and the Kervaire invariant have been known for quite some time \cite{AS,AS2,Zha}. More recently, $\ZM_2$-invariants have played a prominent role in the theory of topological phases. \Red{A review focussing on periodic topological systems and a differential topology perspective on $\ZM_2$-invariants is {\it e.g.} \cite{CMT}. For random topological phases, $\ZM_2$-invariants have to be defined directly as indices \cite{SB,GS,KK,Kel,BCLR}.} For these \Red{random}  systems, it is of interest to \Red{develop} computational tools for $\ZM_2$-invariants. If additional approximate symmetries or conservation laws are present, it is possible to compute the $\ZM_2$-invariant as the parity of an integer invariant that can be obtained from the complex theory (namely a pairing without real symmetries) \cite{Kel,DS}. In situations without such further information, Loring and the second named author have proposed several index formulas for $\ZM_2$-invariants in special cases \cite{Lor,LS1}. 

\vspace{.2cm}

\Red{This article provides a systematic approach to computing $\ZM_2$-valued index pairings} and constructs the so-called {\it skew localizer} for all real index pairings coming from \Red{an unbounded} Fredholm module. It is a real skew-adjoint variant of the spectral localizer of complex index pairings which is known to provide integer-valued pairings via its half-signature \cite{LS1,LS2}. The skew localizer allows to compute the $\ZM_2$-invariant as the sign of its Pfaffian and in $8$ of the $16$ cases as the sign of the determinant of its off-diagonal entry. We believe that the skew localizer is an efficient tool for {\it computational real $K$-theory}, hence playing the same role as the spectral localizer does in complex $K$-theory. \Red{Numerical results will be published elsewhere.}

\vspace{.2cm}

While the first proofs of the connection between (complex) index pairings and spectral localizer were of $K$-theoretic nature \cite{LS1,LS2}, there are now other proofs \cite{LS3,LSS,SS} exploring the tight connection between index pairings and spectral flow (as \Red{formulated} by Phillips \cite{Ph1}). Here all $\ZM_2$-valued real index pairings are shown to be given by suitably defined {\it $\ZM_2$-flows} (various special cases have already been dealt with in \cite{DS2,CPS,DSW}). These flows are then used as the main tool to establish the results on the skew localizer. The new $\ZM_2$-flow pictures of real index pairings are likely of relevance for other applications as well.

\vspace{.2cm}

The version of real index pairings used here is identical to \cite{GS} where it is also explained how index theorems connect them directly to more standard representations of topological invariants, in particular, the strong invariants of topological insulators \cite{PS}. This formulation of real index pairings is tailored to be readily applicable to systems with a symmetry (in particular, topological insulators). As reviewed in Section~\ref{sec-realpairing}, there are $64$ such pairings reflecting that real $K$-theory and real $K$-homology are both $8$-periodic \cite{GVF,Con95}, and $16$ of them take values in $\ZM_2$. It turns out that various $\ZM_2$-flows are associated to these $16$ pairings. One is the $\ZM_2$-valued spectral flow of paths of real skew-adjoint Fredholm operators studied in \cite{CPS}. Below it is slightly generalized to what will be called the orientation flow. Two other $\ZM_2$-flows are the even and odd half-spectral flow for paths of Fredholm operators with a certain reflection symmetry (the odd half-spectral flow was introduced and studied in \cite{DS2}). All these $\ZM_2$-flows are described in Section~\ref{sec-Z2specflow}. They are not independent of each other, it will rather be shown how they can all be reduced to the orientation flow (of a path with extra symmetries, similarly as for parity \cite{DSW}). 

\vspace{.2cm}

Another approach to real index \Red{pairings} is to work with pairings taking values in real Clifford algebras. This is natural from the point of view of Kasparov's $KKO$-theory. In the latter framework, Clifford-valued spectral flows have been studied recently by Bourne, Carey, Lesch and Rennie \cite{BCLR}. There are certainly connections to this work, namely reducing out the Clifford relations should lead to some of the statements of this paper (similar as the Atiyah-Singer classifying spaces \cite{AS} can also be written out more explicitly). However, these connections are not worked out nor used in the present purely functional analytical treatment of $\ZM_2$-valued index \Red{pairings} and flows. In our opinion, this less algebraic approach has the advantage of providing results (as on the skew localizer) that can be implemented directly. 

\vspace{.2cm}

To explain to the reader the main punch line, the remainder of this introduction describes the main results of the paper for one of the $16$ cases and also provides a concrete application to a particular topological insulator (the Kitaev chain). 

\Red{\subsection{Real index pairings}}
\label{eq-RealPairIntro}

Let $\Hh$ be a separable complex Hilbert space,  $P=P^2=P^*$ be an (orthogonal) projection on $\Hh$ and $F=(F^*)^{-1}$ a unitary on $\Hh$ such that the commutator $[P,F]$ is a compact operator. Then $T=PFP+\one-P$ is a Fredholm operator and its index $\Ind(T)=\dim(\Ker(T))-\dim(\Ker(T^*))\in\ZM$ is called the (complex) index pairing of the index pair $(P,F)$. This includes the case where $F=2 E-\one$ is a self-adjoint unitary given in terms of a projection $E$. In this case,  $T=P(2E-\one)P+\one-P $ is still a Fredholm operator which is considered to be the index pairing of an index pair $(P,E)$, given by two projections $P$ and $E$ with compact commutator (note that $(P,E)$ is {\it not} a Fredholm pair in the sense of \cite{Kat,ASS}). In the latter case \Red{of an index pair $(P,E)$}, $T$ is clearly self-adjoint so that $\Ind(T)=0$, but there are reality conditions which may still allow to extract a secondary $\ZM_2$-invariant from $T$. Thus suppose that there is a complex conjugation $\Cc$ on $\Hh$, namely an antilinear isometry squaring to the identity. Given a bounded operator $A\in\Bb(\Hh)$, let $\overline{A}=\Cc A\Cc$ denote its complex conjugate. This allows to formulate real symmetry relations for both $P$ and $E$. In the particular situation considered here, $\overline{E}=E$ is supposed to be real and $P$ to satisfy $\overline{P} =\one-P$. This means that $P$ is even Lagrangian in the terminology \Red{developed in Section~\ref{sec-realpairing}} below. Together the index pair $(P,E)$ is of the type $(j,d)=(2,1)$ in the notation of Section~\ref{sec-realpairing}. In the physical application \Red{in Section~\ref{subsec-Kitaev}}, $P$ is the Fermi projection of a Bogoliobov-de Gennes Hamiltonian in the Majorana representation and $E$ is the Hardy projection associated to the one-dimensional position operator. As already stated above, one has $\Ind(T)=0$, but there is a well-defined $\ZM_2$-invariant
\begin{equation}
\label{eq-def2Ind}
\Ind_2(T)
\;=\;
\dim(\Ker(T))\,\mbox{\rm mod}\;2\;\in\;\ZM_2
\;,
\end{equation}
in the sense that it is a constant under norm-continuous homotopies of $P$ and $E$ respecting the real symmetries imposed as well as the Fredholm property, see \cite{GS}. It is also possible to exchange the roles of $P$ and $E$ and consider the Fredholm operator $T'=E(\one-2P)E+\one-E$. One then finds the same index pairing $\Ind_2(T')=\Ind_2(T)$, see \Red{\eqref{eq-ProjProj}} and \cite{GS}. In  \eqref{eq-def2Ind},  $\ZM_2$ appears as the additive group $\{0,1\}$, but below, {\it e.g.} in \eqref{eq-IntroSf2}, rather as the multiplicative group $\{1,-1\}$. In the following, we freely identify these \Red{groups}.

\Red{\subsection{$\ZM_2$-flow formulas for real index pairings}}

The first main result for the index pairing of $(P,E)$ satisfying the above real symmetries are two $\ZM_2$-flow formulas (stated in Section~\ref{sec-Index=Flow}):
\begin{align}
\Ind_2(T)
&
\;=\;
\SF_2\big(t\in[0,1]\mapsto (1-t)\imath (\one-2P)+t (\one-2E)\imath (\one-2P)(\one-2E)\big)
\label{eq-IntroSf2}
\\
&
\;=\;
\SF\big(t\in[0,\tfrac{1}{2}]\mapsto (1-t)(\one-2E)+t (\one-2P)(\one-2E)(\one-2P)\big)\;\mbox{\rm mod} \;2
\;.
\label{eq-IntroHSf2}
\end{align}
Let us explain the objects \Red{and notions used} in these formulas. First of all, \Red{$\imath=\sqrt{-1}$ denotes the imaginary unit.} In \eqref{eq-IntroSf2} the symmetry of $P$ implies that the operator $J=\imath (\one-2P)$ is real and skew-adjoint, namely $J=\overline{J}=-J^*$. Because $E$ is real, also the operator $(\one-2E)J(\one-2E)$ is real and skew-adjoint. In \eqref{eq-IntroSf2} appears the straight-line path between these two real skew-adjoints which, due to the compactness assumption on $[P,E]$, lies in the real skew-adjoint Fredholm operators. For such paths, the $\ZM_2$-valued spectral flow \Red{$\SF_2$} is well-defined \cite{CPS} and in the present situation of a \Red{linear} path between unitaries given by half of the multiplicity of the eigenvalue $0$ at the midpoint $t=\frac{1}{2}$ of the path modulo 2 \cite{CPS}. For paths of finite-dimensional operators, \Red{the $\ZM_2$-valued spectral flow} is equal to the product of the signs of the Pfaffians of the two endpoint\Red{s} of the path, see Section~\ref{sec-Z2specflow}. The identity \eqref{eq-IntroSf2} already follows from Theorem~8.1 in \cite{CPS}. 

\vspace{.2cm}

To explain \eqref{eq-IntroHSf2}, let us denote the path by $H_t=(1-t)(\one-2E)+t (\one-2P)(\one-2E)(\one-2P)$. For such a path of self-adjoint Fredholm operators, its classical spectral flow \Red{$\SF(t\in[0,1]\mapsto H_t)$} is well-defined \cite{Ph}. However, here one has the reflection property $H_{1-t}=(\one-2P)H_t(\one-2P)$ implying that $\SF(t\in[0,1]\mapsto H_t)$ vanishes. Nevertheless, it will be shown in Section~\ref{sec-EvenHalf} that the parity of the spectral flow on the half-interval $[0,\frac{1}{2}]$ is a well-defined homotopy invariant. That \Red{\eqref{eq-IntroHSf2}} is again equal to $\Ind_2(T)$ is \Red{then} proved in Section~\ref{sec-Index=Flow}. In conclusion, there are two $\ZM_2$-flow representations in this case \Red{corresponding to the two Fredholm operators $T$ and $T'$ described in Section~\ref{eq-RealPairIntro}, namely} one for $\Ind_2(T)$ and one for $\Ind_2(T')$. In fact, \Red{in Section~\ref{sec-Z2specflow}} we provide two such $\ZM_2$-flow representations for all $8$ $\ZM_2$-valued index pairs of either two projections or two unitaries, but only one for index pairs consisting of a projection and a unitary. 

\subsection{\Red{$\ZM_2$-indices via the skew localizer}}

The second main result allows to compute the index pairing \Red{\eqref{eq-def2Ind}} as the sign of the Pfaffian of the skew localizer, provided that $E$ \Red{arises} from an unbounded Fredholm module. Both the even and odd spectral localizer can be used and lead to the same result (see Section~\ref{sec-SkewLoc}). Let $H=H^*=-\overline{H}$ be an invertible such that $P=\chi(H\leq 0)$ \Red{is the spectral projection of $H$ onto $(-\infty,0]$}. \Red{In physical terms,} $H$ is the \Red{Bogoliubov-de Gennes} Hamiltonian in the Majorana representation \Red{and $P$ its Fermi projection}. For example, one can simply choose $H=-\imath J$ with $J=\imath(\one-2P)$, \Red{but working directly with $H$ instead of $P$ in the following is of great practical use because in applications $H$ is given directly while $P$ has to be computed by a full-fledged diagonalization.} Further suppose that there exists a real, self-adjoint and invertible operator  $D_0$ with compact resolvent such that $D_0|D_0|^{-1}=2 E-\one$ and $[D_0,H]$ extends to a bounded operator, notably the operator $D_0$ is given as part of a real \Red{unbounded} Fredholm module used to construct the pairing. Then recall \cite{LS1,LS2} that the odd and even spectral localizers acting on $\Hh\oplus\Hh$ are defined by
$$
L^\odd_{\kappa}
\;=\;
\begin{pmatrix}
\kappa D_0 & H\\
H & - \kappa D_0 
\end{pmatrix}
\;,
\qquad
L^\even_{\kappa}
\;=\;
\begin{pmatrix}
-H & \kappa D_0\\
\kappa D_0 & H
\end{pmatrix}
\;,
$$
where $\kappa>0$ is a tuning parameter. In the odd spectral localizer, the Hamiltonian is used twice to construct a chiral (off-diagonal $2\times 2$) self-adjoint operator, while in the even spectral localizer the off-diagonal entries are self-adjoint and not only adjoints of each other as in the general case (also briefly described in Section~\ref{sec-complexpairing} below). Hence in both cases the complex spectral localizer has supplementary properties (in particular, the normality assumption of \cite{LS2} on the off-diagonal entries of $L^\even_{\kappa}$ is automatically satisfied). Moreover, 
\begin{equation}
\label{eq-PassageOddEven}
L^\odd_{\kappa}
\;=\;
M^*\,L^\even_{\kappa}\,M
\;,
\qquad
M\;=\;
\frac{1}{\sqrt{2}}
\begin{pmatrix}
1 & -1 \\ 1 & 1
\end{pmatrix}
\;,
\end{equation}
so that both contain essentially the same information, reflecting that they are constructed from the same index pair $(P,E)$. In the following let us hence focus on $L^\odd_{\kappa}$ and set
\begin{equation}
\label{eq-DoddR}
D^\odd\;=\;
\begin{pmatrix}
D_0 & 0 \\
0 & -D_0
\end{pmatrix}
\;,
\qquad
R
\;=\;
\frac{1}{2}
\begin{pmatrix}
1-\imath & 1+\imath\\
1+\imath & 1-\imath
\end{pmatrix}
\;.
\end{equation}
The operator $D^\odd$ is real, self-adjoint and invertible, and called the (odd) Dirac operator \Red{(even though it is strictly speaking a doubled Dirac operator)}. Moreover, $R$ is a basis change used to define the skew localizer as
\begin{equation}
\label{eq-SkewLocIntro}
\widehat{L}_\kappa
\;=\;
\imath\,R^*\,L^\odd_\kappa\,R
\;=\;
\begin{pmatrix}
0 & \imath H - \kappa D_0 \\
\imath H + \kappa D_0 & 0
\end{pmatrix}
\;.
\end{equation}
Clearly, this is a real and skew-adjoint operator which in the present case furthermore has the advantage of being off-diagonal (namely, it has a supplementary chiral symmetry). The same holds for 
$$
\widehat{D}
\;=\;
\imath\,R^*\, D^\odd\, R 
\;=\;
\begin{pmatrix}
0 & - D_0 \\
D_0 & 0
\end{pmatrix}
\;.
$$
To state the main result, the skew localizer is now restricted to finite volume. For $\rho>0$ let us set $\Hh_\rho=\Ran(\chi(|D_0|\leq\rho))$ and $(\Hh\oplus\Hh)_\rho=\Hh_\rho\oplus\Hh_\rho$. As $D_0$ has compact resolvent, both $\Hh_\rho$ and $(\Hh\oplus\Hh)_\rho$ are finite dimensional.  The restriction of $\widehat{L}_\kappa$ and $\widehat{D}$ to $(\Hh\oplus\Hh)_\rho$ are denoted by $\widehat{L}_{\kappa,\rho} $ and $\widehat{D}_\rho$. Both are finite-dimensional real skew-adjoint matrices. As such they have well-defined Pfaffians. It is now a fact, essentially already following from \cite{LS1}, that $\widehat{L}_{\kappa,\rho} $ is invertible for $\kappa$ sufficiently small and $\rho$ sufficiently large. In this case the Pfaffian does not vanish and hence has a well-defined sign. The second main result of the paper for the current special case is that
\begin{equation}
\label{eq-IntroMainResPf}
\Ind_2(T)
\;=\;
\sgn\big(\Pf(\widehat{L}_{\kappa,\rho})\big)\,\sgn\big(\Pf(\widehat{D}_\rho)\big)
\;.
\end{equation}
There is another important further simplification in the present case related to the fact that the Pfaffian of an off-diagonal real skew-adjoint is, up to a corrective sign $(-1)^{\frac{1}{2}\dim(\Hh_\rho)(\dim(\Hh_\rho)-1)}$ stemming from the size of the block entries, given by the determinant of the off-diagonal entry. The off-diagonal entries of $\widehat{D}_\rho$ and $\widehat{L}_{\kappa,\rho} $ are given in terms of $D_{0,\rho}$ and $H_\rho$, the finite volume restrictions of $D_0$ and $H$ to $\Hh_\rho$. As the two corrective signs cancel out, 
\begin{equation}
\label{eq-IntroMainRes}
\Ind_2(T)
\;=\;
\sgn(\det(\imath H_\rho + \kappa D_{0,\rho}))\,\sgn(\det(D_{0,\rho}))
\;.
\end{equation}
Let us stress that the numerical tools for the computation of a determinant are much more developed than for a Pfaffian, so that it is better to use \eqref{eq-IntroMainRes} than \eqref{eq-IntroMainResPf} in the present case. Furthermore, in an application one often has given $H$ rather than $P$ which is then costly to obtain by spectral calculus. In \eqref{eq-IntroMainRes} only $H$ itself enters, \Red{as already stressed above}. 

\Red{\subsection{Skew localizer for the Kitaev chain}
\label{subsec-Kitaev}}

As promised above, let us next discuss the (clean) Kitaev chain as an example of a one-dimensional topological insulator to which \eqref{eq-IntroMainRes} can be applied. The Hamiltonian in the Majorana representation acts on $\Hh=\ell^2(\ZM)\otimes \CM^2$ and is given by
$$
H
\;=\;
\imath 
\begin{pmatrix}
0 &   V-\mu \\
-V^*+\mu & 0
\end{pmatrix} 
\;,
$$
where $V$ denotes the right shift on $\ell^2(\ZM)$ and $\mu\in\RM$ the chemical potential. \Red{The complex conjugation on $\ell^2(\ZM)\otimes \CM^2$ is defined element-wise.} Clearly $\imath H$ is a real skew-adjoint which for $\mu\not\in\{-1,1\}$  is invertible. The (Fermi) projection $P=\chi(H\leq 0)$ indeed satisfies $\overline{P}=\one -P$. The off-diagonal entry of the one-dimensional (dual) Dirac operator is given by $D_0=\diag(X,X)$ where $X=\sum_{n\in\ZM}n|n\rangle\langle n|-|0\rangle\langle 0|$ denotes the self-adjoint (unbounded) position operator on $\ell^2(\ZM)$ perturbed at the origin such that it is invertible. The spectral projection $E=\chi(D_0\geq 0)$ is real. It is known ({\it e.g.} \cite{GS}) that the index pairing from $T=P(2E-\one)P+\one-P$ is $\Ind_2(T)=1$ for $|\mu|>1$ and $\Ind_2(T)=-1$ for $|\mu|<1$. Now $\det(D_{0,\rho})=\det(X_\rho)^2>0$ so that by \eqref{eq-IntroMainRes}
\begin{equation}
\label{eq-KitRes}
\Ind_2(T)
\;=\;
\sgn(\det (\imath H_\rho + \kappa D_{0,\rho}))
\;=\;
\sgn
\left(\det
\begin{pmatrix}
\kappa X_\rho &   -V_\rho+\mu \\
V_\rho^*-\mu & \kappa X_\rho 
\end{pmatrix} 
\right)
\;.
\end{equation}
For $\mu=0$ the r.h.s. can be computed analytically with some effort, and one finds that indeed it is $-1$. Clearly for $|\mu|\to\infty$ one has $\Ind_2(T)=1$. Thus the above facts on $\Ind_2(T)$ follow by homotopy invariance for all values of $\mu\not\in\{-1,1\}$. Let us further note that \eqref{eq-KitRes} is precisely Loring's guess in Section 4.3 of \cite{Lor} (at least for $\kappa=1$ and without the finite-volume restriction). In this work, we provide a criterion under which the operator $\imath H_\rho +\kappa D_{0,\rho}$ is invertible for suitable choices of $\kappa$ and $\rho$ as well as a proof that it indeed provides the desired $\ZM_2$-index.

\Red{\subsection{Structure of the article}}

Let us conclude this introduction with a brief outline of the sections below. Section~\ref{sec-complexpairing} reviews known results about complex index pairings, their spectral flow interpretation \cite{Ph1} and expression in terms of the spectral localizer \cite{LS1, LS2, LS3, LSS}. Section~\ref{sec-realpairing}, still a review, presents real index pairings by following closely \cite{GS}. In Section~\ref{sec-Z2specflow}, the various notions of $\ZM_2$-flows and their basic properties are discussed, in particular, variants of the identities \eqref{eq-IntroSf2} and \eqref{eq-IntroHSf2} are proved. Section~\ref{sec-SkewLoc} then deals with the skew localizer. \Red{Section~\ref{sec-SkewLoc}} provides a more constructive approach to the basis change $R$ that was introduced \Red{in \eqref{eq-DoddR}} in an ad-hoc manner. The main new results generalize equation \eqref{eq-IntroMainResPf} to all $16$ real index pairings with values in $\ZM_2$. \Red{In particular, these results} prove corrected versions of the conjectures in Section~1.7 of \cite{LS1} which concerned $4$ of the $16$ cases. For $8$ of these $\ZM_2$-values pairings (the upper $\ZM_2$-diagonal in the Kitaev table recalled in Theorem~\ref{theo-indexlist} below), the skew localizer has a supplementary chiral symmetry allowing to bring it into an off-diagonal form as in \eqref{eq-SkewLocIntro}, which implies that variations of \eqref{eq-IntroMainRes} hold for these $8$ cases. In this work, no further applications of the results to topological insulators are given, even though this is relatively immediate given the prior works \cite{PS,GS}. This together with a numerical implementation will be the object of a future work.

\vspace{.2cm}

\noindent {\bf Acknowledgement:} This work was partially supported by the DFG grant SCHU 1358/6-2.

\vspace{.2cm}

\section{Review of complex index pairings}
\label{sec-complexpairing}

The complex index pairing of a projection $P$ with a unitary $F$ having a compact commutator $[P, F]$ is given by the Fredholm operator 
\begin{equation}
\label{eq-BasicFred}
T
\;=\;
PFP+\one-P
\end{equation}
and its index $\Ind(T)=\dim(\Ker(T))-\dim(\Ker(T^*))$.  It can be computed as a spectral flow of a norm-continuous path $t\mapsto H_t$ of self-adjoint Fredholm operators in the sense of Phillips \cite{Ph}. We choose conventions such that $s\mapsto \SF(t\in[0,s]\mapsto H_t)$ is right-continuous.

\begin{theo}[\cite{Ph1}, see also \cite{DS2}]
\label{theo-complexindex}
Let $t\in[0,1]\mapsto H_t$ be a norm-continuous path of self-adjoint operators with invertible endpoints $H_0$ and $H_1$ such that $H_t-H_0$ is compact for all $t\in [0,1]$ and $H_1=F^*H_0F$. If $P=\chi(H_0\leq0)$, then 
$$
\Ind(PFP+\one-P)\;=\;\SF(t\in[0,1]\mapsto H_t)\;.
$$
In particular, one has for the linear path connecting $\one-2P$ and $F^*(\one-2P)F$ 
$$
\Ind(PFP+\one-P)
\;=\;
\SF\big(t\in[0,1]\mapsto (1-t)(\one-2P)+tF^*(\one-2P)F\big)\;.
$$
\end{theo}

If the index \Red{pairing} results from a pairing between a $K$-theory class and an unbounded Fredholm module (also called spectral triple), it can be computed in terms of the spectral localizer. Two cases have to be distinguished: either $P$ is a projection specifying a $K_0$-class in some $C^*$-algebra and $F$ is the Dirac phase of the Dirac operator $D^\even$ making up an even unbounded Fredholm module for this algebra, or $F$ is a unitary specifying a $K_1$-class in some $C^*$-algebra and $P$ is the Hardy projection of the Dirac operator $D^\odd$ of an odd unbounded Fredholm module for this algebra. To distinguish these cases also notationally, in the latter case the projection will be denoted by $E$ and the unitary by $U$.  To further enlarge the flexibility of the results, let us suppose in the even case that $P=\chi(H\leq 0)$ is the negative spectral projection of an invertible self-adjoint operator $H$ (in physical terms, $P$ is the Fermi projection of the gapped Hamiltonian $H$). In the odd case, let $U=A|A|^{-1}$ be the phase of an invertible operator $A$ and then set 
\begin{equation}
\label{eq-chiralHam}
H
\;=\;
\begin{pmatrix}
0 & A \\ A^* & 0
\end{pmatrix}
\end{equation}
(in physical terms, $U$ is the Fermi unitary of the gapped chiral Hamiltonian $H$).

\vspace{.2cm}

Let us first describe the spectral localizer in the even case \cite{LS2, LSS,SS}. The self-adjoint, invertible even Dirac operator $D^\even$ on $\Hh \oplus \Hh$  is of the form
$$
D^\even
\;=\;
\begin{pmatrix}
0 & D_0^*\\
D_0 & 0
\end{pmatrix}\;,
$$
where $D_0$ is an invertible, unbounded operator on $\Hh$.  It is supposed that \Red{$D_0$} has a compact resolvent and that the commutator $[D^\even, H\oplus H]$ extends to a bounded operator. Note that the even Dirac operator is odd (or chiral) in the sense that $\Gamma D \Gamma=-D$ for $\Gamma=\diag(\one_\Hh, -\one_\Hh)$. \Red{In the standard terminology of $K$-homology \cite{HR,GVF}, $(D,\Gamma)$ specifies an even unbounded Fredholm module (also called an even spectral triple) for a smooth subalgebra of the $C^*$-algebra generated by $H$.} As stated above, the Dirac phase is now denoted $F=D_0|D_0|^{-1}$. It is then a standard result that $T$ given by \eqref{eq-BasicFred} is indeed a Fredholm operator \cite{GVF}. As in \cite{LS2, LSS}, we will always assume that $D_0$ is normal. While this \Red{assumption} can be circumvented \cite{SS} (see also the remark at the end of Section~\ref{sec-j2d0}), it holds in all cases of interest below and is thus sufficient. The even spectral localizer is defined as the operator
\begin{equation}
 \label{eq-evenLoc}
L^\even_{\kappa}
\;=\;
\begin{pmatrix}
-H & \kappa D_0^*\\
\kappa D_0 & H
\end{pmatrix}
\;,
\end{equation}
acting on $\Hh \oplus \Hh$ where $\kappa>0$ is a tuning parameter. To construct finite volume restrictions of the spectral localizer, let us now set
$$
\Hh_\rho\;=\;\Ran(\chi(|D_0|\leq \rho))\;,
\qquad
(\Hh\oplus \Hh)_\rho\;=\;\Ran(\chi(|D^\even|\leq \rho))\;,
$$
for a radius $\rho >0$. Recall that $D^\even$ has compact resolvent so that each $\Hh_\rho$ and $(\Hh\oplus\Hh)_\rho$ is finite-dimensional.  As $D_0$ is normal, one has $(\Hh\oplus \Hh)_\rho=\Hh_\rho\oplus\Hh_\rho$.  \Red{Let $\pi_\rho: \Hh \to \Hh_\rho$, denote the surjective partial isometry onto $\Hh_\rho$ with $\Ker(\pi_\rho)=(\Hh_\rho)^\perp$ and such that  $\pi_\rho|_{\Hh_\rho}$ is the identitiy on $\Hh_\rho$. By abuse of notations, the surjective partial isometry onto $(\Hh\oplus \Hh)_\rho$ is also denoted by $\pi_\rho: \Hh\oplus\Hh \to (\Hh\oplus\Hh)_\rho$.}  Also let $\one_\rho=\pi_\rho\pi_\rho^*$ be the identity on $\Hh_\rho$ and on $(\Hh\oplus \Hh)_\rho$. For any operator $B$ on $\Hh$ or $\Hh\oplus \Hh$, we set $B_\rho=\pi_\rho B \pi_\rho^*$ which is an operator on $\Hh_\rho$ or $(\Hh\oplus \Hh)_\rho$.  With these notations, the finite volume spectral localizer on $(\Hh\oplus \Hh)_\rho$ is
$$
L^\even_ {\kappa,\rho}
\;=\;
\begin{pmatrix}
-H_\rho & \kappa D_{0,\rho}^*\\
 \kappa D_{0,\rho} & H_\rho
\end{pmatrix}
\;.
$$
%

\begin{theo}[\cite{LS2,LSS}]
\label{tho-LSS}
Let $g=\Vert H^{-1}\Vert^{-1}$ be the gap of the invertible self-adjoint operator $H$. Suppose that  
\begin{equation}
\label{eq-rhoevenSpecLocd0j2}
\kappa\;\leq\; \frac{g^3}{12\Vert H \Vert\, \Vert[D_0,H]\Vert}\;,
\qquad
\rho\;>\;\frac{2g}{\kappa}\;.
\end{equation}
Then $(L^\even_ {\kappa,\rho})^2\geq\frac{g^2}{4}\one_\rho$. In particular, $L^\even_ {\kappa,\rho}$ is invertible and thus has a well-defined signature $\Sig(L^\even_ {\kappa,\rho})$ which is independent of $\kappa$ and $\rho$ \Red{satisfying \eqref{eq-rhoevenSpecLocd0j2}}, and
$$
\Ind(PFP+\one-P)\;=\;\frac{1}{2}\,\Sig(L^\even_ {\kappa,\rho})\;.
$$
\end{theo}

Now let us consider the odd case. The self-adjoint and invertible Dirac operator is supposed to be of the form
$$
D^\odd
\;=\;
\begin{pmatrix} 
D_0 & 0 \\ 0 & -D_0
\end{pmatrix}
\;,
$$
have a compact resolvent and be such that the commutator $[A,D_0]$ extends to a bounded operator. \Red{In the standard terminology of $K$-homology \cite{HR,GVF}, $D_0$ specifies an odd unbounded Fredholm module (or odd spectral triple) for a smooth subalgebra of the $C^*$-algebra generated by $A$. In \cite{HR,GVF}, $D_0$ is then called the Dirac operator, while here we call also $D^\odd$ the Dirac operator as this slight deviation of standard terminology allows to treat the even and odd cases with real symmetries simultaneously in a convenient manner.} Let $E=\chi(D_0\geq 0)$. Then $T=EUE+\one-E$ with $U= A|A|^{-1}$ is a Fredholm operator giving the pairing of $U$ with $E$. The odd spectral localizer is defined as the operator
\begin{equation}
\label{eq-oddLoc}
L_\kappa^\odd\;=\;\begin{pmatrix}
\kappa D_0 & A\\
A^* & -\kappa D_0
\end{pmatrix}\;,
\end{equation}
acting on $\Hh \oplus \Hh$ where $\kappa>0$ is a tuning parameter. The partial isometries $\pi_\rho$ and finite volume restrictions are defined exactly as in the even case. The finite volume odd spectral localizer on $\Hh_\rho \oplus \Hh_\rho$ is defined by
$$
L^\odd_ {\kappa,\rho}\;=\;\begin{pmatrix}
\kappa D_ {0,\rho} & A_\rho\\
A_\rho^* & -\kappa D_{0,\rho}
\end{pmatrix}\;.
$$
%

\begin{theo}[\cite{LS3}]
\label{tho-LS2}
Let $g=\Vert A^{-1}\Vert^{-1}$ be the gap of the invertible operator $A$. Suppose that
\begin{equation}
\label{eq-kappa0odd}
\kappa\;<\; \frac{g^3}{12\Vert A \Vert \Vert[D_0,A]\Vert}\;,
\qquad
\frac{2g}{\kappa}\;<\;\rho\;.
\end{equation}
Then the matrix $L^\odd_ {\kappa,\rho}$ satisfies the bound $(L^\odd_ {\kappa,\rho})^2\geq\frac{g^2}{4}\one_\rho$. In particular, $L^\odd_ {\kappa,\rho}$ is invertible and thus has a well-defined signature $\Sig(L^\odd_ {\kappa,\rho})$. It is independent of $\kappa$ and $\rho$ \Red{satisfying \eqref{eq-kappa0odd}}, and
$$
\Ind(EUE+\one-E)
\;=\;\frac{1}{2}\,\Sig(L^\odd_ {\kappa,\rho})
\;.
$$
\end{theo}

\section{Review of real index pairings}
\label{sec-realpairing}

To define real symmetry properties, suppose given a complex conjugation $\Cc:\Hh\to\Hh$ on the Hilbert space $\Hh$. This is an anti-linear isometric map squaring to the identity. As above, the complex conjugate of an operator $A$ on $\Hh$ is $\overline{A}=\Cc A\Cc$, and its transpose $A\Red{^T}=(\overline{A})^*$ is the adjoint of the complex conjugate. Following \cite{DS2,GS}, it will be convenient to introduce some terminology.

\begin{defini}
\label{def-SymmetryOp} 
A symmetry operator on a Hilbert space $\Hh$ with complex conjugation $\Cc$ is a real unitary $S=\overline{S}=(S^*)^{-1}\in\Bb(\Hh)$ squaring to either the identity or minus the identity. In these respective cases, \Red{$S$} is called even or odd. Let $A$ be an operator on $\Hh$ and $P$ a projection \Red{on $\Hh$}.

\begin{enumerate}[{\rm (i)}]

\item $A$ is called even real if $S^*\overline{A}S=A$ for some even symmetry $S$. 

\item $A$ is called odd real if $S^*\overline{A}S=A$ for some odd symmetry $S$. 

\item $A$ is called even symmetric if $S^*A\Red{^T}S=A$ for some even symmetry $S$.  

\item $A$ is called odd symmetric if $S^*A\Red{^T}S=A$ for some odd symmetry $S$.  

\item  $P$ is called even Lagrangian if $S^*\overline{P}S=\one-P$ for some even symmetry $S$. 

\item  $P$ is called odd Lagrangian if $S^*\overline{P}S=\one-P$ for some odd symmetry $S$. 

\end{enumerate}

\end{defini}

Note that $S=\one$ is a possible choice for an even symmetry. Thus a real operator is even real  \Red{w.r.t. $S=\one$}. \Red{It is also natural to address an  odd real operator as quaternionic.} Furthermore, \Red{an operator} $A$ is odd symmetric if and only if $SA$ is anti-symmetric. Moreover, $P$ is even/odd Lagrangian  w.r.t. $S$ if and only if $\imath(\one-2P)$ is even/odd real w.r.t. $S$. Finally, Lagrangian planes in symplectic geometry are maximally isotropic subspaces w.r.t. the symplectic form which is an odd symmetry, so that the associated projections are odd Lagrangian in the above terminology.

\vspace{.2cm}

Now let us consider index pairings $T=PFP+\one-P$ as in \eqref{eq-BasicFred} and suppose that $P$ and $F$ have real symmetries in the sense of Definition~\ref{def-SymmetryOp}. One possibility is that these symmetries do not restrict the value of the index of \Red{$T$}, another one that \Red{the index} is always even, and yet another one that \Red{the index} vanishes. In the latter case, it may be possible to define a secondary invariant using $\Ind_2(T)=\dim(\Ker(T))\,\mbox{\rm mod}\,2$, whenever it is well-defined in the sense that it is a constant under norm-continuous homotopies of  $P$ and $F$ respecting the symmetries imposed and the Fredholm property of the index pairing. All these possibilities appear in Theorem~\ref{theo-indexlist} below. It can, moreover, be of interest to consider index pairings of two projections with real symmetries (see \Red{the example in} the introduction) and also of two unitaries with real symmetries. For the former case, let be given two projections $P$ and $E$ with compact commutator $[P,E]$, then the operators $2E-\one$ and $\one-2P$ are both unitary, one thus has two index pairings
\begin{equation}
\label{eq-ProjProj}
T\;=\;P(2\,E\,-\,\one)P\,+\,\one-P
\;,
\qquad
T'\;=\;E(\one\,-\,2\,P)E\,+\,\one-E
\;,
\end{equation}
of the type \eqref{eq-BasicFred}. Both have vanishing index $\Ind(T)=\Ind(T')=0$ because both $T$ and $T'$ are self-adjoint, but it is possible that $\Ind_2(T)$ and $\Ind_2(T')$ are well-defined and non-trivial. It is then an elementary result that $\Ind_2(T)=\Ind_2(T')$ (Proposition 1 in \cite{GS}). Similarly, given two unitaries $U$ and $F$ having a compact commutator $[U,F]$, one obtains two projections on the doubled Hilbert space $\Hh \oplus \Hh$ by setting $P=\frac{1}{2}\binom{\one\;\;-U}{-U^*\;\;\one}$ and $E=\frac{1}{2}\binom{\one\;\;F}{F^*\;\;\one}$. Consequently, there are again two index pairings of the form \eqref{eq-BasicFred}:
\begin{equation}
\label{eq-UniUni}
T\;=\;P\begin{pmatrix} F & 0 \\ 0 & F\end{pmatrix} P\,+\,\one-P
\;,
\qquad
T'\;=\;E\begin{pmatrix} U & 0 \\ 0 & U\end{pmatrix} E\,+\,\one-E
\;.
\end{equation}
By Proposition~2 in \cite{GS},  one has again $\Ind(T)=\Ind(T')=0$ and $\Ind_2(T)=\Ind_2(T')$. Hence pairings of two projections and two unitaries may lead to $\ZM_2$-indices only. The following result from \cite{GS} specifies which type of index is well-defined and non-trivial for given real symmetries.

\begin{theo}
\label{theo-indexlist}
Let $P$ and $E$ be projections and $U$ and $F$ be untaries such that the commutators $[P,E]$, $[P,F]$, $[U,E]$ and $[U,F]$ are compact.  \Red{Further let $S$ and $\Sigma$ be commuting symmetries such that $[S,F]=[S,E]=0$ and $[\Sigma, P]=[\Sigma, U]=0$. Then let
$d\in\{0,1,\ldots,7\}$ specify the symmetry properties of $E$ and $F$ as given in second row of the table below, and let 
$j\in\{0,1,\ldots,7\}$ specify the symmetry properties of $P$ and $U$ as given in second column of that table. Then for each given pair $d,j\in\{0,1,\ldots,7\}$ specifying $P$ $E$, $U$ and $F$ as well as their symmetry properties, let us consider just as above the operator
\begin{equation}
\label{eq-pairingdef}
T
\;=\;
\left\{
\begin{array}{cc}
E\,UE\,+\,\one-E\;, &    d\; \mbox{\rm odd and }j\;\mbox{\rm odd}\;,
\\
E\,(\one-2P)\,E\,+\,\one-E\;, &   d\; \mbox{\rm odd and }j\;\mbox{\rm even}\;,
\\
PFP\,+\one-P\;, &     d\; \mbox{\rm even and }j\; \mbox{\rm even}\;,
\\
\frac{1}{2}\begin{pmatrix} \one & -U \\ -U^* & \one\end{pmatrix}\begin{pmatrix} F & 0 \\ 0 & F\end{pmatrix} \frac{1}{2}\begin{pmatrix} \one & -U \\ -U^* & \one\end{pmatrix}\,+\,\one-\frac{1}{2}\begin{pmatrix} \one & -U \\ -U^* & \one\end{pmatrix}\;, &d\; \mbox{\rm even and }j\; \mbox{\rm odd}\;,
\end{array}
\right.
\end{equation}
In all cases, $T$ is a Fredholm operator and  the index pairing is of the type indicated in the following table:
\begin{center}
\begin{tabular}{|c|c||c|c|c|c|c|c|c|c|}
\hline
 &  & $d=0$ & $d=1$ & $d=2$ & $d=3$ & $d=4$ & $d=5$ & $d=6$ & $d=7$
\\
\hline
\multirow{2}{*}{$j$} &  & ${\scriptstyle\Sigma^*\overline{F}\Sigma=F}$ & ${\scriptstyle\Sigma^*\overline{E}\Sigma=E}$ & ${\scriptstyle\Sigma^*F\Red{^T}\Sigma=F}$ & ${\scriptstyle\Sigma^*\overline{E}\Sigma=\one-E}$ & ${\scriptstyle\Sigma^* \overline{F}\Sigma=F}$ & ${\scriptstyle\Sigma^*\overline{E}\Sigma=E}$ & ${\scriptstyle\Sigma^*F\Red{^T}\Sigma=F}$ & ${\scriptstyle\Sigma^*\overline{E}\Sigma=\one-E}$
\\
 & & ${\scriptstyle\Sigma^2=\one}$  & ${\scriptstyle\Sigma^2=\one}$  & ${\scriptstyle\Sigma^2=\one}$ & ${\scriptstyle\Sigma^2=-\one}$ & ${\scriptstyle\Sigma^2=-\one}$  & ${\scriptstyle\Sigma^2=-\one}$ & ${\scriptstyle\Sigma^2=-\one}$ & ${\scriptstyle\Sigma^2=\one}$
\\
\hline
\hline
\multirow{2}{*}{$0$} & ${\scriptstyle S^*\overline{P}S=P}$ & \multirow{2}{*}{$\ZM$} & \multirow{2}{*}{$0$} & \multirow{2}{*}{$0$} & \multirow{2}{*}{$0$} & \multirow{2}{*}{$2\,\ZM$} & \multirow{2}{*}{$0$} & \multirow{2}{*}{$\ZM_2$} & \multirow{2}{*}{$\ZM_2$}
\\
& ${\scriptstyle S^2=\one}$
& &  &   & & & & &
\\
\hline
\multirow{2}{*}{$1$} & ${\scriptstyle S^*\overline{U}S=U}$ 
& \multirow{2}{*}{$\ZM_2$} & \multirow{2}{*}{$\ZM$} & \multirow{2}{*}{$0$} & \multirow{2}{*}{$0$} & \multirow{2}{*}{$0$}   & \multirow{2}{*}{$2 \,\ZM$}  &  \multirow{2}{*}{$0$} &  \multirow{2}{*}{$\ZM_2$}
\\
&  ${\scriptstyle S^2=\one}$
& &  &   & & & & &
\\
\hline
\multirow{2}{*}{$2$} & ${\scriptstyle S^*\overline{P}S=\one-P}$  & \multirow{2}{*}{$\ZM_2$} & \multirow{2}{*}{$\ZM_2$} & \multirow{2}{*}{$\ZM$} & \multirow{2}{*}{$0$} & \multirow{2}{*}{$0$} & \multirow{2}{*}{$0$} & \multirow{2}{*}{$2\,\ZM$} & \multirow{2}{*}{$0$}
\\
&  ${\scriptstyle S^2=\one}$
& &  &   & & & & &
\\
\hline
\multirow{2}{*}{$3$} & ${\scriptstyle S^*U\Red{^T}S=U}$ 
& \multirow{2}{*}{$0$} &  \multirow{2}{*}{$\ZM_2$} & \multirow{2}{*}{$\ZM_2$} & \multirow{2}{*}{$\ZM$} & \multirow{2}{*}{$0$} & \multirow{2}{*}{$0$} & \multirow{2}{*}{$0$} & \multirow{2}{*}{$2\,\ZM$}
\\
&  ${\scriptstyle S^2=-\one}$
& &  &  & & & & &
\\
\hline
\multirow{2}{*}{$4$} & ${\scriptstyle S^*\overline{P}S=P}$  & \multirow{2}{*}{$2\,\ZM$} & \multirow{2}{*}{$0$} & \multirow{2}{*}{$\ZM_2$} & \multirow{2}{*}{$\ZM_2$} & \multirow{2}{*}{$\ZM$} & \multirow{2}{*}{$0$} & \multirow{2}{*}{$0$} & \multirow{2}{*}{$0$}
\\
&  ${\scriptstyle S^2=-\one}$
& &  &  & & & & &
\\
\hline
\multirow{2}{*}{$5$} & ${\scriptstyle S^*\overline{U}S=U}$ 
& \multirow{2}{*}{$0$} & \multirow{2}{*}{$2\,\ZM$} & \multirow{2}{*}{$0$} & \multirow{2}{*}{$\ZM_2$} & \multirow{2}{*}{$\ZM_2$} & \multirow{2}{*}{$\ZM$} & \multirow{2}{*}{$0$} & \multirow{2}{*}{$0$}
\\
&  ${\scriptstyle S^2=-\one}$
& &  &   & & & & &
\\
\hline
\multirow{2}{*}{$6$} & ${\scriptstyle S^*\overline{P}S=\one-P}$   & \multirow{2}{*}{$0$} & \multirow{2}{*}{$0$} & \multirow{2}{*}{$2\,\ZM$} & \multirow{2}{*}{$0$} & \multirow{2}{*}{$\ZM_2$} & \multirow{2}{*}{$\ZM_2$} & \multirow{2}{*}{$\ZM$} & \multirow{2}{*}{$0$}
\\
&  ${\scriptstyle S^2=-\one}$
& &  &   & & & & &
\\
\hline
\multirow{2}{*}{$7$} & $ {\scriptstyle S^*U\Red{^T}S=U}$ 
& \multirow{2}{*}{$0$} & \multirow{2}{*}{$0$} & \multirow{2}{*}{$0$} & \multirow{2}{*}{$2\,\ZM$} & \multirow{2}{*}{$0$} & \multirow{2}{*}{$\ZM_2$} & \multirow{2}{*}{$\ZM_2$} & \multirow{2}{*}{$\ZM$}
\\
 & ${\scriptstyle S^2=\one}$
 & &  &   & & & & &
\\
\hline
\end{tabular}
\end{center}
For each entry $\ZM$, the symmetires do not restrict the value of the index of $T$. For each entry $2\,\ZM$, the index of $T$ is always even. For each entry $\ZM_2$ one has $\Ind(T)=0$ and the $\ZM_2$-index of $T$, defined by \eqref{eq-def2Ind}, is constant under norm-continuous homotopies of $P$, $E$, $U$ and $F$ respecting the symmetries imposed and the Fredholm property of $T$.
For each entry $0$, one has $\Ind(T)=0$ and, moreover, there exists a path of index pairings with the given symmetries in an augmented Hilbert space which connects the pairing to a pairing with trivial $\ZM_2$-index.}
\end{theo}

\Red{To help the reader extract information from this lengthy statement of  Theorem~\ref{theo-indexlist}, let us consider explicitly the case $(j,d)=(1,0)$. Then $S$ and $\Sigma$ are even symmetry operators such that $[S,\Sigma]=0$ and $U$ and $F$ are unitaries such that the commutator $[U,F]$ is compact and such that $S^*\overline{U}S=U$ and $\Sigma^*\overline{F}\Sigma=F$. Moreover, it is assumed that $F$ commutes with $S$ and $U$ commutes with $\Sigma$. Then for $T$ as given by the fourth row of \eqref{eq-pairingdef} the usual $\ZM$-valued index vanishes. But according to Theorem~\ref{theo-indexlist} the $\ZM_2$-index $\Ind_2(T)\in\ZM_2$ of $T$ is well-defined and it is a homotopy invariant under deformations respecting all symmetries and Fredholm properties.
}

\vspace{.2cm}

When $P$ and $U$ are $K$-theoretic data in an algebra with an anti-linear involution $A\mapsto \overline{A}$, the symmetry relations in Theorem~\ref{theo-indexlist} correspond to $KR$-classes. Similarly, when $E$ and $F$ result from an even or odd Fredholm module respectively, the symmetries specify $KR$-homology classes \cite{Con95,GVF} (\Red{none of these facts are used below}).

\section{$\ZM_2$-flows and associated index formulas}
\label{sec-Z2specflow}

In this section the various $\ZM_2$-flows and their basic properties are described and then index formulas are proved connecting them to index pairs. This leads to a $\ZM_2$-flow interpretation for all $\ZM_2$-indices of the table in Theorem~\ref{theo-indexlist}. Let us note that Section~\ref{subsec-Z2specflow} on the $\ZM_2$-valued spectral flow is merely a review of \cite{CPS}, while Section~\ref{sec-Oflow} is new. Section~\ref{sec-EvenHalf} on the even half-spectral flow is new, but actually nothing but the even \Red{counterpart} of the odd half-spectral flow of Section~\ref{sec-OddHSF} which was already studied in \cite{DS2}. The proofs in Section~\ref{sec-EvenHalf} are, however, much simpler than those in \cite{DS2} and transpose to Section~\ref{sec-OddHSF}. Section~\ref{sec-Index=Flow} then contains the $\ZM_2$-flow formulas for all $\ZM_2$-valued index pairings that were already advertised in the abstract and \Red{the} introduction. These formulas are relatively direct consequences of Sections~\ref{sec-Oflow} to \ref{sec-OddHSF}.

\subsection{$\ZM_2$-valued spectral flow}
\label{subsec-Z2specflow}

In \cite{CPS} a $\ZM_2$-valued spectral flow for paths of real skew-adjoint Fredholm operators on a Hilbert space $\Hh$ with complex conjugation is studied. Let us begin with a finite dimensional $\Hh$ and consider a path $t\in [0,1]\mapsto T_t$ of skew-adjoint real matrices with invertible endpoints. \Red{The} $\ZM_2$-valued spectral flow \Red{of this path} is defined as the sign of the product of the Pfaffians of the endpoints of the path 
\begin{equation}
\label{eq-SFPF}
\SF_2(t\in [0,1]\mapsto T_t)\;=\;\sgn(\Pf(T_0))\,\sgn(\Pf(T_1))\;.
\end{equation}
Actually, in \cite{CPS} one finds $\SF_2(t\in [0,1]\mapsto T_t)=\sgn(\det(A))$ where $A=\overline{A}$ is such that $T_1=A\Red{^T}T_0A$, but taking the Pfaffian of the latter relation readily leads to \eqref{eq-SFPF}. It is obvious that the definition \eqref{eq-SFPF} is a homotopy invariant (under homotopies of paths with invertible endpoints) and has a concatenation property. Furthermore, the $\ZM_2$-valued spectral flow is additive under direct sums (in the additive group $\ZM_2$). \Red{In the complex case, these are precisely the properties used to characterize the spectral flow \cite{Les}. Moreover, if} the path $t\in [0,1] \mapsto T_t$ is real analytic, all eigenvalues can be chosen to be analytic (also at eigenvalue crossings). As the spectrum of \Red{skew-adjoints} has the reflection symmetry $\sigma(T_t)=-\sigma(T_t)$, the finite set of points at which the invertibility breaks down are such eigenvalue crossings through $0$. By Theorem~4.4 in \cite{CPS} the $\ZM_2$-valued spectral flow of a real analytic path equals the sum of all eigenvalue crossings along the path, each one counted with its multiplicity, modulo 2. Now clearly there is no \Red{spectrum} flowing at each of these eigenvalue crossing, there is rather a change of orientation between the eigenbasis \Red{from the left to the right} of the crossing. Therefore, it appears to us more precise and adequate to use the term {\it orientation flow} instead of $\ZM_2$-valued spectral flow. We will use an admittedly slight generalization to justify a change of terminology in the next section. 

\vspace{.2cm}

The extension of the finite dimensional definition to a path $t\in [0,1]\mapsto T_t$ of real skew-adjoint Fredholm operators on an infinite dimensional Hilbert space $\Hh$ with invertible endpoints $T_0$ and $T_1$ can be carried out exactly as in Phillips' treatment of spectral flow \cite{Ph}.  The details are given in \cite{CPS}. From now on we will use that $\SF_2(t\in [0,1]\mapsto T_t)$ is well-defined and invariant under continuous homotopies of paths of real skew-adjoint Fredholm operators leaving the endpoints fixed (Theorem~4.3 in \cite{CPS}). It remains additive under direct sums.

\vspace{.2cm}

Reference \cite{CPS} also contains a first index formula for the $\ZM_2$-valued spectral flow. For a real skew-adjoint unitary $J=\overline{J}=-J^*=-J^{-1}$ and a real unitary $F=\overline{F}$ such that $[J,F]$ is compact, the $\ZM_2$-valued spectral flow of the linear path $t\in [0,1]\mapsto (1-t)J+tFJF^*$ connecting $J$ to $FJF^*$ is well-defined. By Theorem~8.1 in \cite{CPS}, one has \eqref{eq-IntroSf2}, namely for $P=\chi(\imath J\leq0)$ 
\begin{equation}
\label{eq-IndexFormulaCPS}
\SF_2(t\in [0,1]\mapsto (1-t)J+tFJF^*)\;=\;\Ind_2(PFP+\one-P)
\;.
\end{equation}
%

\subsection{Orientation flow}
\label{sec-Oflow}

The $\ZM_2$-valued spectral flow can be slightly generalized. Let $\SymProd $ be an even symmetry operator and $t\in [0,1]\mapsto T_t$ a continuous path of Fredholm-operators with invertible endpoints such that $T_t=\SymProd^*\overline{T_t}\SymProd =-T_t^*$ holds for all $t \in [0,1]$. As $\SymProd $ is self-adjoint, its unitary square root $\Rr$ with spectrum $\sigma(\Rr)=\{1,\imath\}$ is well-defined and $t\in [0,1]\mapsto \Rr^*T_t\Rr$ is a path of real skew-adjoints with invertible endpoints. Let us define the orientation flow of the path $t\in [0,1]\mapsto T_t$ as
\begin{equation}
\label{eq-OfDef}
\Of_2\big(t\in [0,1]\mapsto T_t\big)
\;=\;
\SF_2\big(t\in [0,1]\mapsto \Rr^*T_t\Rr\big)\;.
\end{equation}
If $\Hh$ is finite dimensional and therefore $t\in [0,1]\mapsto \Rr^* T_t\Rr $ is a path of real skew-adjoint matrices, one has
$$
\Of_2\big(t\in [0,1]\mapsto T_t\big)\;=\;\sgn(\Pf(\Rr^*T_0\Rr))\,\sgn(\Pf(\Rr^*T_1\Rr))\;.
$$
The orientation flow is invariant under continuous homotopies of paths of skew-adjoint Fredholm operators that are even real w.r.t. $\SymProd $,  leaving the endpoints fixed. Note that $\sigma(T_t)=\sigma(\Rr^*T_t\Rr)$. Thus if the path $t \mapsto T_t$  is real analytic, its orientation flow equals the sum of all eigenvalue crossings along the path, each one counted with its multiplicity, modulo 2. Furthermore, the orientation flow inherits the additivity under direct sums.

\vspace{.2cm}

Let us now generalize the index formula \eqref{eq-IndexFormulaCPS} in two directions: the unitary $F$ is only supposed to be even real and the end points of the path are not necessarily unitary, but only invertible. Hence let $P$ be a projection such that $\SymProd^*\overline{P}\SymProd =\one-P$ and $F=\SymProd^*\overline{F}\SymProd $ be a unitary such that $[P,F]$ is compact. Let $\thetaSFeven(P,F,\SymProd )$ denote the set of paths $t \in [0,1]\mapsto T_t=\SymProd^*\overline{T_t}\SymProd =-T_t^*$ of skew-adjoint Fredholm operators such that 
\begin{enumerate}	
\item[{\rm (i)}] 
$T_0$ is invertible\;,
\item[{\rm (ii)}] 
$P=\chi(\imath T_0\leq 0)$\;,
\item[{\rm (iii)}]
$T_t - T_0$ is compact for all $t \in [0, 1]$\;,
\item[{\rm (iv)}] 
$T_1=FT_0F^*$\;.
\end{enumerate}
%

\begin{theo}
\label{indextheoSF}
For all paths $t \in [0,1]\mapsto T_t$ in $\thetaSFeven(P,F,\SymProd )$
$$
\Of_2(P,F)\;=\;\Of_2(t\in [0,1]\mapsto T_t)
$$
is independent of the path and
$$
\Ind_2(PFP+\one-P)
\;=\;
\Of_2(P,F)
\;=\;
\Of_2(P,F^*)\;.
$$
\end{theo}

\noindent {\bf Proof.} Let $t \in [0,1]\mapsto T_t$ be a path in $\thetaSFeven(P,F,\SymProd )$. The path $t \in [0,1]\mapsto \Rr^*T_t\Rr$ connects $\Rr^*T_0\Rr$ to 
$$
\Rr^*T_1\Rr
\;=\; 
 \Rr^* FT_0F^*\Rr
\;=\;
 (\Rr^* F\Rr) (\Rr^* T_0\Rr) (\Rr^* F\Rr)^*\;.
$$
As $F=\SymProd^*\overline{F}\SymProd $, $\Rr^* F\Rr$ is a real unitary.
As $T_0-T_1$ is compact, there is a homotopy within the paths of real skew-adjoint Fredholm operators leaving the endpoints invariant, connecting the path $t \in [0, 1] \mapsto \Rr^*T_t\Rr$ to the linear path 
$$
t\in [0,1]\mapsto (1-t)\Rr^*T_0\Rr+t(\Rr^* F\Rr) (\Rr^* T_0\Rr) (\Rr^* F\Rr)^*
$$ 
connecting $\Rr^*T_0\Rr$ to $\Rr^*T_1\Rr$. Thus, recalling the definition \eqref{eq-OfDef}, 
\begin{align*}
\Of_2(t \in [0, 1] \mapsto  \Red{T_t})
\;=\;
\SF_2
\big(
t\in [0,1]\mapsto (1-t)\Rr^*T_0\Rr+t(\Rr^* F\Rr) (\Rr^* T_0\Rr) (\Rr^* F\Rr)^*
\big)
\;,
\end{align*}
by Theorem~4.\Red{3} in \cite {CPS}. As
\begin{align*}
(t,s)\in [0,1]\times [0,1] \mapsto 
(1-t)\Rr^*T_0|T_0|^{-s}\Rr+t(\Rr^* F\Rr) (\Rr^* T_0|T_0|^{-s}\Rr) (\Rr^* F\Rr)^*
\end{align*}
is a homotopy of real skew-adjoint Fredholm operators, 
\begin{align*}
&\SF_2
\big( t\in[0,1]\mapsto(1-t)\Rr^*T_0\Rr+t(\Rr^* F\Rr) (\Rr^* T_0\Rr) (\Rr^* F^*\Rr)
\big)
\\
&\;=\;\SF_2\big(s\in[0,1]\mapsto \Rr^*T_0|T_0|^{-s}\Rr\big)\\
&\;\;\;\;\;\;+\;\SF_2\big( t\in[0,1]\mapsto(1-t)\Rr^*T_0|T_0|^{-1}\Rr+t(\Rr^* F\Rr) (\Rr^* T_0|T_0|^{-1}\Rr) (\Rr^* F\Rr)^*\big)\\
&\;\;\;\;\;\;+\;\SF_2\big(s\in[0,1]\mapsto(\Rr^* F\Rr) (\Rr^* T_0|T_0|^{-1+s}\Rr) (\Rr^* F\Rr)^*\big)
\\
&\;=\;\SF_2\big( t\in[0,1]\mapsto(1-t)\Rr^*T_0|T_0|^{-1}\Rr+t(\Rr^* F\Rr) (\Rr^* T_0|T_0|^{-1}\Rr) (\Rr^* F\Rr)^*\big)\;,
\end{align*}
because $s\in[0,1]\mapsto \Rr^*T_0|T_0|^{-s}\Rr$ and $s\in[0,1]\mapsto(\Rr^* F\Rr) (\Rr^* T_0|T_0|^{-1+s}\Rr) (\Rr^* F\Rr)^*$ are paths of invertibles and therefore their $\ZM_2$-valued spectral flows vanish. Now $\Rr^*T_0|T_0|^{-1}\Rr$ is a real skew-adjoint unitary and $\Rr^* F\Rr$ is a real unitary so that Theorem~8.1 in \cite{CPS} applies. As $\chi(\imath \Rr^*T_0|T_0|^{-1}\Rr\leq0)=\Rr^*P\Rr\;,$
\begin{align*}
\Ind_2(PFP+\one-P)
& \;=\;\Ind_2\big((\Rr^*P\Rr)(\Rr^* F\Rr)(\Rr^*P\Rr)+\one-(\Rr^*P\Rr)\big)\\
&\;=\;\SF_2\big( t\in[0,1]\mapsto(1-t)\Rr^*T_0\Rr+t(\Rr^* F\Rr) (\Rr^* T_0\Rr) (\Rr^* F\Rr)^*\big)\;,
\end{align*}
concluding the proof of the first equality. The second one holds because $\Ind_2(PFP+\one-P)=\Ind_2(PF^*P+\one-P)$.
\hfill $\Box$

\subsection{Even half-spectral flow}
\label{sec-EvenHalf}

The even half-spectral flow for paths of self-adjoint Fredholm operators with even symmetric endpoints is defined as follows. Let $\SymProd \in \Bb(\Hh)$ be an even symmetry operator and $F=-\SymProd^*F\Red{^T}\SymProd $ unitary. For an even real projection $P=\SymProd^*\overline{P}\SymProd $ such that $[P,F]$ is compact, let $\thetaHSFeven(P,F,\SymProd )$ denote the set of continuous paths $t \in [0,1]\mapsto H_t$ of self-adjoint Fredholm operators such that 
\begin{enumerate}	
\item[{\rm (i)}] 
$H_0=\SymProd^*(H_0)\Red{^T}\SymProd$ is invertible,
\item[{\rm (ii)}] 
$P=\chi(H_0\leq 0)$\;,
\item[{\rm (iii)}]
$H_t - H_0$ is compact for all $t \in [0, 1]$\;,
\item[{\rm (iv)}] 
$H_{1-t}=( \SymProd  F)^\ast (H_t)\Red{^T} ( \SymProd  F)$\;.
\end{enumerate}
Note that (iv) implies $H_1=F^\ast H_0 F$. Combined with (iii) this implies that $[H_0, F]$ is compact. Furthermore,  $\sigma(H_{1-t})=\sigma(H_t)$. Hence the spectral flow of the path $t \in [0, 1] \mapsto H_t$ vanishes.

\begin{lemma}
\label{lem-evenHSF}
For a self-adjoint invertible operator $H=H^\ast=\SymProd^* H\Red{^T}\SymProd $ and a unitary operator $F=-\SymProd^\ast F\Red{^T} \SymProd $ such that the commutator $[H, F]$ is compact, the linear path $t \in [0, 1] \mapsto H_t = H +t F^\ast [H, F]$ lies in $\theta_\even(P, F, \SymProd )$ for $P=\chi(H\leq0)$.
\end{lemma}

\noindent {\bf Proof.} As 
\begin{align*}
( \SymProd  F)^\ast (H_t)\Red{^T} (\SymProd  F) 
&\; = \; 
F^\ast ((1-t)H+ t \SymProd^* F\Red{^T}H\Red{^T}\overline{F} \SymProd )F\\
&\; = \; 
(1-t)F^\ast HF+t F^\ast F \SymProd^* H\Red{^T}  \SymProd  F^\ast F\\
&\; = \; (1-t)F^\ast HF+t H 
\; = \; H_{1-t}\; ,
\end{align*}
and (i), (ii) and (iii) hold by assumption, the claim follows.
\hfill $\Box$

\begin{lemma}
\label{lem-Kramer}
Every finitely degenerate eigenvalue of $ H_\frac{1}{2}$  has even multiplicity.
\end{lemma}

\noindent {\bf Proof.} One has $F^\ast \SymProd^\ast\overline{H_\frac{1}{2}} \SymProd  F=H_\frac{1}{2}$ or equivalently $H_\frac{1}{2}F^\ast  \SymProd^\ast=F^* \SymProd^\ast\overline{H_\frac{1}{2}}$. If $\lambda\in \RM$ is an eigenvalue of $H_\frac{1}{2}$ of finite multiplicity, $H_\frac{1}{2}v=\lambda v$ holds for some $v \in \Hh\setminus \{0\}$. Then $H_\frac{1}{2}F^\ast \SymProd^\ast\overline{v}=F^\ast \SymProd^\ast\overline{H_\frac{1}{2}}\overline{v}=\lambda F^\ast \SymProd^\ast\overline{v}$ follows, where $\overline{v}=\mathcal{C}v$ is the complex conjugate of $v$. Let us show that $v$ and $F^\ast \SymProd^\ast\overline{v}$ are linearly independent. Suppose the contrary, namely there is some ${\rm a} \in \CM \setminus \{0\}$ such that ${\rm a}v =F^\ast \SymProd^\ast \overline{v}$ or equivalently $v=\frac{1}{{\rm a}}F^\ast \SymProd^\ast\overline{v}$ holds.  Then $v= \frac{1}{{\rm a}}F^\ast \SymProd^\ast\overline{\frac{1}{{\rm a}}F^\ast \SymProd^\ast\overline{v}}=\frac{1}{|{\rm a}|^2}F^\ast \SymProd^\ast F\Red{^T} \SymProd^\ast v=-\frac{1}{|{\rm a}|^2}F^\ast F v=-\frac{1}{|{\rm a}|^2}v$, which implies $v=0$. This shows that $v$ and $F^\ast \SymProd^\ast\overline{v}$ are linearly independent and $\lambda$ is at least twice degenerate.  If the multiplicity of $\lambda$ is larger than two, the same argument applies to the orthogonal complement of ${\rm span}\{v,F^\ast \SymProd^\ast\overline{v}\}$. Repeating this procedure a finite number of times shows the claim.
\hfill $\Box$

\begin{proposi}
\label{prop-Z2flowwellef}
For $P$ and $F$ as in {\rm Lemma~\ref{lem-evenHSF}},
$$
\HSF(P,F)
\;=\;
\SF\big(t \in [0, \tfrac{1}{2}] \mapsto H_t\big)
\;\mbox{\rm mod}\;2\;\in\ZM_2
$$
is well-defined and independent of the choice of the path $t \in [0, 1]\mapsto H_t$ in $\theta_\even(P, F, \SymProd )$ and thus defines a $\ZM_2$-flow called the {\rm even half-spectral flow}. In particular, it can be calculated by 
$$
\SF\big(t \in [0, \tfrac{1}{2}] \mapsto \one-2P+t F^*[\one-2P, F]\big)
\;\mbox{\rm mod}\;2
\; .
$$
\end{proposi}

\noindent {\bf Proof.} Let $t \in [0, 1] \mapsto H_t$  be a path in $\theta_{\even}(P, F, \SymProd )$ and let us set $ H_t'=\one-2P+t F^*[\one-2P, F]$. Then, as $H|H|^{-1}=\one-2P$, $(t, s) \in [0,1]\times[0,1] \mapsto H(t, s)=sH'_t+(1-s)H_t$ is a homotopy of paths in $\theta_{\even}(P, F, \SymProd )$ such that $H(t,0)=H_t$ and $H(t,1)=H_t'$. By the homotopy invariance of the spectral flow, one has 
\begin{align*}
&\SF(t \in [0, \tfrac{1}{2}]\mapsto H(t, 0) )+\SF(s \in [0,1] \mapsto H(\tfrac{1}{2}, s)) \\
&-\SF(t \in [0, \tfrac{1}{2}]\mapsto H(t, 1) )-\SF(s \in [0,1] \mapsto H(0, s))
\;= \;0\;.
\end{align*}
By Lemma~\ref{lem-Kramer}, $\SF(s \in [0,1] \mapsto H(\frac{1}{2},s))$ is even. As $H(0,s)$ is invertible for all $s \in [0,1]$,  $\SF(s \in [0,1] \mapsto H(0, s))=0$, thus 
$$
\SF(t \in [0, \tfrac{1}{2}]\mapsto H(t, 0) )
\;=\;
\SF((t \in [0, \tfrac{1}{2}]\mapsto H(t, 1) ))
\;\mbox{\rm mod}\;2
\; ,
$$
concluding the proof. 
\hfill $\Box$

\vspace{.2cm}

Let us note that by choice of convention, the spectral flow is right-continuous. However, if one works with a left-continuous spectral flow, the definition of the half-spectral flow remains unchanged due to Lemma~\ref{lem-Kramer}. 

 \begin{proposi}
\label{theo-locconst} 
On the set
\begin{align*}
\PM(\mathcal{H},  \SymProd )\;=\;\{(H, F):H=H^*= \SymProd^\ast \overline{H} \SymProd  \;{\rm invertible},\; F=\left(F^*\right)^{-1}=- \SymProd^\ast F\Red{^T}\SymProd ,\; [H, F] \;{\rm compact}\}
\end{align*}
the even half-spectral flow $(H,F)\in \PM(\mathcal{H},  \SymProd )\mapsto\HSF(P,F)$ is locally constant.
\end{proposi}

\noindent {\bf Proof.} Let $s \in [0, 1] \mapsto (H(s), F(s)) \in \PM(\mathcal{H},  \SymProd )$ be a continuous path and set $H(t, s)=H(s)+t F(s)^*[H(s), F(s)]$. For all $s \in [0, 1]$, the half-spectral flow $\SF(t \in [0, \tfrac{1}{2}] \mapsto H(t, s))\,\mbox{\rm mod}\,2$ is defined. For $s, s' \in [0, 1]$,
\begin{align*}
&\SF(t \in [0, \tfrac{1}{2}]\mapsto H(t, s) )\;+\;\SF(\tilde{s} \in [s, s'] \mapsto H(\tfrac{1}{2}, \tilde{s})) \\
&\;\;-\;\SF((t \in [0, \tfrac{1}{2}]\mapsto H(t, s') ))\;-\;\SF(\tilde{s} \in [s, s'] \mapsto H(0, \tilde{s}))
\;= \;0\;.
\end{align*}
By Lemma~\ref{lem-Kramer}, $\SF(\tilde{s} \in [s, s'] \mapsto H(\frac{1}{2},\tilde{s}))$ is even. Moreover, $\SF(\tilde{s} \in [s, s'] \mapsto H(0, \tilde{s}))$ vanishes as $ H(0, \tilde{s})$ is invertible for all $\tilde{s} \in [s, s']$. Thus $\SF(t \in [0, \frac{1}{2}]\mapsto H(t, s) )=\SF((t \in [0, \frac{1}{2}]\mapsto H(t, s') )) \,\mbox{\rm mod}\,2$.
\hfill $\Box$

\vspace{.2cm}

Now all is set up for the proof of the index formula for the even half-spectral flow.

\begin{theo}
\label{theo-indexflow}
For $(H, F) \in \PM(\mathcal{H}, \SymProd )$ and $P=\chi (H \leq 0)$, one has
$$
\Ind_2(PFP+\one-P)
\;=\;
\HSF(P,F)
\;=\;
\HSF(P,F^*)
\; .
$$
\end{theo}

\noindent {\bf Proof.} By Proposition~\ref{prop-Z2flowwellef} it is sufficient to consider the linear path
$$
t\in[0,1]\mapsto H_t
\;=\;
(1-t)(\one-2P)+tF^*(\one-2P)F\;.
$$ 
As $H_t$ is invertible for $t \neq \frac{1}{2}$ one has $\SF(t \in [0, \frac{1}{2}]\mapsto H_t)=\SF(t \in [\frac{1}{2}-\epsilon, \frac{1}{2}]\mapsto H_t)\,\mbox{\rm mod}\,2 $ for some $\epsilon>0$. For $\epsilon$ sufficiently small and $a>0$ such that $a \notin \sigma(H_t)$ for $t \in [\frac{1}{2}-\epsilon, \frac{1}{2}]$ and $\dim(\Ran(\chi_{[-a, a]}(H_{\frac{1}{2}-\epsilon})))=\dim(\Ker(H_\frac{1}{2}))$, one has 
$$
\SF(t \in [\tfrac{1}{2}-\epsilon, \tfrac{1}{2}]\mapsto H_t)
\;=\;\dim(\Ker(H_\frac{1}{2}))-\dim(\Ran(\chi_{[0, a]}(H_{\frac{1}{2}-\epsilon})))
\;,
$$
so that by Lemma~\ref{lem-Kramer}
$$
\SF(t \in [\tfrac{1}{2}-\epsilon, \tfrac{1}{2}]\mapsto H_t)\;\mbox{\rm mod } 2
\;=\;\dim(\Ran(\chi_{[0, a]}(H_{\frac{1}{2}-\epsilon})))\;\mbox{\rm mod } 2\;.
$$
As $H_{\frac{1}{2}-\epsilon}$ is conjugate to $\overline{H_{\frac{1}{2}+\epsilon}}$, one has $\dim( \Ran(\chi_{[0, a]}(H_{\frac{1}{2}-\epsilon})))=\dim(\Ran(\chi_{[0, a]}(H_{\frac{1}{2}+\epsilon})))$. Now as all eigenvalues of $H_t$ contributing to the kernel of $H_\frac{1}{2}$ change their sign at $t=\frac{1}{2}$, it follows that $\dim(\Ran(\chi_{[0, a]}(H_{\frac{1}{2}+\epsilon})))=\dim(\Ran(\chi_{[-a, 0]}(H_{\frac{1}{2}-\epsilon})))$. This implies $\dim(\Ran(\chi_{[0, a]}(H_{\frac{1}{2}-\epsilon})))=\dim(\Ran(\chi_{[-a, 0]}(H_{\frac{1}{2}-\epsilon})))$. Thus $\dim(\Ran(\chi_{[0, a]}(H_{\frac{1}{2}-\epsilon})))=\frac{1}{2}\dim(\Ker(H_\frac{1}{2}))$ and  
$$
\SF(t \in [0, \tfrac{1}{2}]\mapsto H_t)\;=\;\tfrac{1}{2}\dim(\Ker(H_\frac{1}{2}))
\;\mbox{\rm mod}\;2\;.
$$ 
One has $H_\frac{1}{2}=\frac{1}{2}\Red{\one}-P+\frac{1}{2}\Red{\one}-F^\ast PF=\one-P-F^\ast PF$. For some vector $v=v_1+v_2$ in the kernel of $H_\frac{1}{2}$ such that $Pv_1=v_1$ and $Pv_2=0$,  one has
\begin{align*}
(\one-P-F^\ast PF)v\;=\;0\;
&\Longleftrightarrow \; v_1+v_2-v_1-F^\ast PFv_1-F^\ast PFv_2\;=\;0\\
&\Longleftrightarrow\; F^\ast PFv_1+F^\ast PFv_2\;=\;v_2\\
&\Longleftrightarrow\; PFv_1+PFv_2\;=\;Fv_2\\
&\Longleftrightarrow\; PFPv_1\;=\;(\one-P)F(\one-P)v_2
\;.
\end{align*}
Hence $0=PFPv_1=
(PFP+\one-P)v_1$ and $0=(\one-P)F(\one-P)v_2=
\big((\one-P)F(\one-P)-P\big)v_2$. Thus 
$$
\Ker(H_\frac{1}{2})
\;=\;
\Ker(PFP+\one-P)\oplus \Ker((\one-P)F(\one-P)+P)
\;.
$$ 
By Lemma 1 in \cite{GS}
\begin{align*}
\dim(\Ker((\one-P)F(\one-P)+P))\;&=\;\dim(\Ker(PF^*P+\one-P))\\
&=\;\dim(\Ker(\overline{P}F\Red{^T}\overline{P}+\one-\overline{P}))
\\
&=\;\dim(\Ker(PFP+\one-P))\; ,
\end{align*}
where the last equality follows from $ \SymProd^\ast(\overline{P}F\Red{^T}\overline{P}+\one-\overline{P})\SymProd =-PFP+\one-P$. This implies the first equality claimed. The second follows from $\Ind_2(PFP+\one-P)=\Ind_2(PF^*P+\one-P)$.
\hfill $\Box$

\vspace{.2cm}

There is an interpretation of the even half-spectral flow as orientation flow (and likely vice versa, but this is not investigated here). For any path $t \in [0,1]\mapsto H_t \in \theta_{\even}(P, F, \SymProd )$, the kernel of $H_\frac{1}{2}$ is even dimensional by Lemma~\ref{lem-evenHSF}. By adding a compact perturbation $K$ on the kernel of $H_\frac{1}{2}$ such that $K=( \SymProd  F)^\ast K\Red{^T} ( \SymProd  F)$ one may assume that $H_\frac{1}{2}$ is invertible. In particular, the projection onto $\Ker(H_\frac{1}{2})$ is such a perturbation. Then set
\begin{equation}
\label{eq-T0T1}
T_0\,=\,
\imath \begin{pmatrix}
H_0 & 0\\
0 & -H_0
\end{pmatrix}
\;,
\qquad
T_1\;=\;\imath \begin{pmatrix}
H_\frac{1}{2} & 0\\
0 & -\SymProd^* \overline{H_\frac{1}{2}}\SymProd 
\end{pmatrix}
\;,
\qquad
Q\;=\;\begin{pmatrix}
0& \SymProd \\
\SymProd^*  & 0
\end{pmatrix}\;.
\end{equation}
Both $T_0$ and $T_1$ are skew-adjoint and even real w.r.t. $Q$. As $T_0-T_1$ is compact, the orientation flow of the linear path connecting $T_0$ to $T_1$ is well-defined. 

\begin{proposi}
\label{prop-evenHSF}
Let $t \in [0,1]\mapsto T_t$ be a path of skew-adjoint operators connecting $T_0$ and $T_1$ defined in \eqref{eq-T0T1} and such that $T_t-T_0$ is compact for all $t$ and $T_t=Q^*\overline{T_t}Q$. Then
$$
\HSF(P,F)
\;=\;
\Of_2(t \in [0, 1] \mapsto T_t)\,.
$$
\end{proposi}

\noindent {\bf Proof.} By the homotopy invariance of the orientation flow it is sufficient to show
$$
\HSF(P,F)
\;=\;
\Of_2(t \in [0, 1] \mapsto (1-t)T_0+tT_1)\,.
$$
By Proposition~\ref{prop-Z2flowwellef},
$$
\HSF(P,F)
\;=\;
\SF\big(t \in [0, 1] \mapsto (1-t)H_0+tH_\frac{1}{2}\big)
\;\mbox{\rm mod}\;2\;.
$$
On the other hand, one has $\sigma((1-t)T_0+tT_1)=\sigma(\imath((1-t)H_0+tH_\frac{1}{2}))\cup\big(-\sigma(\imath((1-t)H_0+tH_\frac{1}{2}))\big)$,
because $(1-t)H_0+tH_\frac{1}{2}$ and $(1-t)H_0+t\SymProd^*\overline{H_\frac{1}{2}}\SymProd $ are isospectral due to the relation $\SymProd^*\overline{((1-t)H_0+tH_\frac{1}{2})}\SymProd =(1-t)H_0+t\SymProd^*\overline{H_\frac{1}{2}}\SymProd $. Therefore, as the paths are analytic so that the orientation flow just counts the number of eigenvalue crossings modulo $2$,
$$
\Of_2\big(t \in [0, 1] \mapsto (1-t)T_0+tT_1\big)
\;=\;\SF\big(t \in [0, 1] \mapsto (1-t)H_0+tH_\frac{1}{2}\big)
\;\mbox{\rm mod}\;2\;.
$$
This implies the claim.
\hfill $\Box$

\subsection{Odd half-spectral flow}
\label{sec-OddHSF}

In \cite{DS2} a half-spectral flow for paths of self-adjoint Fredholm operators with odd symmetric endpoints conjugate to each other by an odd symmetric unitary is studied. More precisely, let $\SymProd\in \Bb(\Hh)$ be an odd symmetry operator and $F=\SymProd^*F\Red{^T}\SymProd$ an odd symmetric unitary. For an odd real projection $P=\SymProd^*\overline{P}\SymProd$ such that $[P,F]$ is compact, let $\thetaHSFodd(P,F,\SymProd)$ denote the set of continuous paths $t \in [0,1]\mapsto H_t$ of self-adjoint Fredholm operators such that 
\begin{enumerate}	
\item[{\rm (i)}] 
$H_0=\SymProd^*(H_0)\Red{^T}\SymProd$ is invertible\;,
\item[{\rm (ii)}] 
$P=\chi(H_0\leq 0)$\;,
\item[{\rm (iii)}]
$H_t - H_0$ is compact for all $t \in [0, 1]$\;,
\item[{\rm (iv)}] 
$H_{1-t}=( \SymProd F)^\ast (H_t)\Red{^T} ( \SymProd F)$\;.
\end{enumerate}
By Proposition~5.2 in \cite{DS2}
$$
\SF(t \in [0,\tfrac{1}{2}]\mapsto H_t)\;\mbox{\rm mod}\;2
$$
is well-defined for any path $t \in [0,1]\mapsto H_t \in \thetaHSFodd(P,F,\SymProd)$. Strictly speaking, \cite{DS2} only considers unitary $H_0$, but \Red{$\SF(t \in [0,\tfrac{1}{2}]\mapsto H_t)$ mod $2$ is well-defined for any path $t \in [0,1]\mapsto H_t \in \thetaHSFodd(P,F,\SymProd)$} by the argument \Red{leading to Proposition~\ref{prop-Z2flowwellef}} in Section~\ref{sec-EvenHalf}. We will refer to it here as the odd half-spectral flow of the path $t \in [0,1]\mapsto H_t$. As $s\in[0,1] \mapsto H_0|H_0|^{-s}$ is a paths of self-adjoint invertibles connecting $H_0$ to $\one-2P$ and $H_0|H_0|^{-s}=\SymProd^*(H_0|H_0|^{-s})\Red{^T}\SymProd$ holds for all $s\in[0,1]$ and $[H_0|H_0|^{-s},F]$ is compact  for all $s\in[0,1]$ by Proposition~5.4 in \cite{DS2}, the quantity
$$
\HSF(P,F)
\;=\;
\SF(t \in [0,\tfrac{1}{2}]\mapsto H_t)\;\mbox{\rm mod}\;2
$$
is independent of the path $t \in [0,1]\mapsto H_t \in \thetaHSFodd(P,F,\SymProd)$.

\vspace{.2cm}

The proof of the following theorem, already stated as Theorem~6.1 in \cite{DS2},  can be carried out along the lines of the proof of Theorem~\ref{theo-indexflow} (which is  a considerable improvement over the argument in \cite{DS2}).

\begin{theo}
\label{indextheoHSFodd}
For $P$ and $F$ as above, one has
$$
\Ind_2(PFP+\one-P)
\;=\;\HSF(P,F)
\;=\;\HSF(P,F^*)
\;.
$$
\end{theo}

As above there is an interpretation of the odd half-spectral flow as $\ZM_2$-valued spectral flow. For any path $t \in [0,1]\mapsto H_t \in \thetaHSFodd(P,F,\SymProd)$ the kernel of $H_\frac{1}{2}$ is even dimensional. By adding a compact perturbation $K$ on the kernel of $H_\frac{1}{2}$ such that $K=( \SymProd F)^\ast K\Red{^T} ( \SymProd F)$ one may assume that $H_\frac{1}{2}$ is invertible. In particular, the projection onto ${\Ker(H_\frac{1}{2})}$ is such a perturbation. Then define $T_0$, $T_1$ and $Q$ by \eqref{eq-T0T1}. Still $T_0$ and $T_1$ are skew-adjoint and even real w.r.t. $Q$. As $T_0-T_1$ is compact, the orientation flow of the linear path connecting $T_0$ to $T_1$ is well-defined. The proof of the following result is essentially the same as that of Proposition~\ref{prop-evenHSF}.

\begin{proposi}
\label{prop-oddhlaf}
Let $t \in [0,1]\mapsto T_t$ be a path of skew-adjoint operators connecting $T_0$ and $T_1$ given by \eqref{eq-T0T1} such that $T_t-T_0$ is compact for all $t$ and such that $T_t=Q^*\overline{T_t}Q$. Then
$$
\HSF(P,F)
\;=\;
\Of_2(t \in [0, 1] \mapsto T_t)\,.
$$
\end{proposi}

\subsection{$\ZM_2$-valued real index pairings as $\ZM_2$-flows}
\label{sec-Index=Flow}

In this section, a $\ZM_2$-flow interpretation of all $\ZM_2$-indices of the table in Theorem~\ref{theo-indexlist} is given. All of them can be thought of as $\ZM_2$-versions of Theorem~\ref{theo-complexindex}. It is somewhat lengthy to spell out all the details, but this is needed as a preparation for the next section. We will proceed case by case, always regrouping two at a time.

\vspace{.2cm}

For $(j,d)\in\{(1,7),(5,3)\}$, one considers the index pairing $T=EUE+\one-E$. In these cases $\SymProd=\Sigma S$ is an even symmetry operator and $E$ is even Lagrangian and $U$ is even real w.r.t. $\SymProd$.  Therefore Theorem~\ref{indextheoSF} directly implies the following:

\begin{coro}
\label{coro-index=flowd3j5}
For $(j,d)\in\{(1,7),(5,3)\}$ one has
$$
\Ind_2(EUE+\one-E)
\;=\;
\Of_2(E,U)
\;.
$$
In particular,
\begin{align*}
\Ind_2(EUE+\one-E)
&\;=\;
\Of_2(t\in[0,1]\mapsto (1-t)\imath (2E-\one)+tU\imath (2E-\one)U^*)\\
&\;=\;
\Of_2(t\in[0,1]\mapsto (1-t)\imath(\one-2E)+tU\imath(\one-2E)U^*)\;.
\end{align*}
\end{coro}

For $(j,d)\in\{(3,1),(7,5)\}$, one considers the index pairing $T=EUE+\one-E$. Then $\SymProd=\Sigma S$ is an odd symmetry operator and $\one-2E$ and $U$ are odd symmetric w.r.t. $\SymProd$. Theorem~\ref{indextheoHSFodd} thus implies

\begin{coro}
\label{coro-index=flowd1j3}
For $(j,d)\in\{(3,1),(7,5)\}$ one has
$$
\Ind_2(EUE+\one-E)
\;=\;
\HSF(E,U)
\;.
$$
In particular, for the linear path $t\in[0,1]\mapsto (1-t)(\one-2E)+tU^*(\one-2E)U$ one has
$$
\Ind_2(EUE+\one-E)
\;=\;
\SF(t \in [0,\tfrac{1}{2}]\mapsto (1-t)(\one-2E)+tU^*(\one-2E)U)\;\mbox{\rm mod}\;2\;.
$$
\end{coro}

Also in the following corollaries one can write out explicit formulas using the straight-line paths just as in Corollaries~\ref{coro-index=flowd3j5} and \ref{coro-index=flowd1j3}. This is not spelled out, however. Next let us consider the case $(j,d)\in\{(2,0),(6,4)\}$ and hence an index pairing $T=PFP+\one-P$.  Then $\SymProd=\Sigma S$ is an even symmetry operator and $P$ is even Lagrangian and $F$ is even real w.r.t. $\SymProd$.  Theorem~\ref{indextheoSF} implies

\begin{coro}
\label{coro-index=Z2flowd0j2}
For $(j,d)\in\{(2,0),(6,4)\}$ one has
$$
\Ind_2(PFP+\one-P)
\;=\;
\Of_2(P,F)
\;.
$$
\end{coro}

For $(j,d)\in\{(0,6),(4,2)\}$ the index pairing is again $T=PFP+\one-P$. Now $\SymProd=\Sigma S$ is an odd symmetry operator and $\one-2P$ and $F$ are odd symmetric w.r.t. $\SymProd$. From Theorem~\ref{indextheoHSFodd} follows

\begin{coro}
\label{coro-indes=halfflowd2j4}
For $(j,d)\in\{(0,6),(4,2)\}$ one has
$$
\Ind_2(PFP+\one-P)
\;=\;
\HSF(P,F)
\;.
$$
\end{coro}

For $(j,d)\in\{(1,0),(5,4)\}$, the index pairing is
$T=P\binom{F\;\;0}{0\;\;F}P+\one-P$ with $P=\frac{1}{2}\binom{\one\;\;-U}{-U^*\;\;\one}$. Moreover,  $P$ is even Lagrangian and $\diag(F,F)$ is even real  w.r.t. the even symmetry operator $\SymProd=\diag(\Sigma S, -\Sigma S)$. Furthermore $E=\frac{1}{2}\binom{\one\;\;F}{F^*\;\;\one}$ as above is even Lagrangian and $\diag(U,U)$  is even real w.r.t. $\SymProd$. Now Theorem~\ref{indextheoSF} implies the following

\begin{coro}
\label{prop-FPd0j1}
For $(j,d)\in\{(1,0),(5,4)\}$ one has
$$
\Ind_2(P\diag(F,F)P+\one-P)
\;=\;
\Of_2(P,\diag(F,F))
\;=\;
\Of_2(E,\diag(U,U))
\;.
$$
\end{coro}

For $(j,d)\in\{(3,2),(7,6)\}$, the index pairing is again $T=P\binom{F\;\;0}{0\;\;F}P+\one-P$ with $P=\frac{1}{2}\binom{\one\;\;-U}{-U^*\;\;\one}$, but now $\one-2P$ and $\diag(F,F)$ are odd symmetric w.r.t. $\SymProd=\binom{0\;\;\Sigma S}{\Sigma S\;\;0}$. Moreover, for $E=\frac{1}{2}\binom{\one\;\;F}{F^*\;\;\one}$ the unitaries $(\one-2E)$ and $\diag(U,U)$  are odd symmetric w.r.t. $\SymProd$. Thus by Theorem~\ref{indextheoHSFodd}

\begin{coro}
\label{coro-index=halfflowd2j3}
For $(j,d)\in\{(3,2),(7,6)\}$ one has
$$
\Ind_2(P\diag(F,F)P+\one-P)
\;=\;
\HSF(P,\diag(F,F))
\;=\;
\HSF(E,\diag(U,U))
\;.
$$
\end{coro}

For $(j,d)\in\{(0,7),(4,3)\}$ one considers the index pairing $T=E(\one-2P)E+\one-E$. Then $\SymProd=\Sigma S$ is an even symmetry operator and $E$ is even Lagrangian and $P$ is even real w.r.t. $\SymProd$. By Theorem~\ref{theo-indexflow} and Theorem~\ref{indextheoSF}

\begin{coro}
\label{coro-index=halfflowd3j4}
For $(j,d)\in\{(0,7),(4,3)\}$ one has
$$
\Ind_2(E(\one-2P)E+\one-E)
\;=\;
\HSF(P,\one-2E)
\;=\;
\Of_2(E, \one-2P)\;.
$$
\end{coro}

For $(j,d)\in\{(2,1),(6,5)\}$, the index pairing is $T=E(\one-2P)E+\one-E$. Moreover, $\SymProd=\Sigma S$ is an even symmetry operator and $P$ is even Lagrangian and $E$ is even real w.r.t. $\SymProd$. As in Corollary~\ref{coro-index=halfflowd3j4} one concludes:

\begin{coro}
\label{coro-index=halfflowd1j2}
For $(j,d)\in\{(2,1),(6,5)\}$,
$$
\Ind_2(E(\one-2P)E+\one-E)
\;=\;
\HSF(E,\one-2P)
\;=\;
\Of_2(P,\one-2E)
\;.
$$
\end{coro}

\section{Skew localizer for real index pairings}
\label{sec-SkewLoc}

In this section, the real index pairing is supposed to result from \Red{an unbounded} Fredholm module for either $P=\chi(H\leq0)$ or $U=A|A|^{-1}$ as described in Section~\ref{sec-complexpairing}. Hence \Red{one is} given either an odd or even Dirac operator $D$ (we suppress the upper index of Section~\ref{sec-complexpairing}) and $E=\chi(D_0\geq 0)$ or $F=D_0|D_0|^{-1}$, respectively. It is, moreover, assumed that the real symmetry relations of $E$ and $F$ are inherited from the Dirac operator which hence satisfies $\Sigma^* \overline{D_0}\Sigma=\pm D_0^{(*)}$ (this means that $D_0$ specifies an unbounded representative of a $KR$-homology class \Red{for the smooth algebras generated by $H$ or $A$ respectively}, see \cite{Con95,HR,GVF,GS}). Furthermore, $[S,D_0]=0$. Associated to indices $(j,d)$, the index pairing is constructed as in \eqref{eq-pairingdef}  and the operators entering it are supposed to satisfy the symmetry relations stated in the table of Theorem~\ref{theo-indexlist} which involve the commuting symmetry operators $S$ and $\Sigma$. The focus is only on those cases $(j,d)$ which lead to a $\ZM_2$-entry in that table. The symmetries and other operators will often be expressed in terms of the standard Pauli matrices
$$
\sigma_0
\;=\;
\begin{pmatrix}
\one & 0 \\ 0 & \one
\end{pmatrix}
\;,
\qquad
\sigma_1
\;=\;
\begin{pmatrix}
0 & \one  \\ \one & 0
\end{pmatrix}
\;,
\qquad
\sigma_2
\;=\;
\begin{pmatrix}
0 & -\imath\, \one  \\ \imath\,\one & 0
\end{pmatrix}
\;,
\qquad
\sigma_3
\;=\;
\begin{pmatrix}
\one & 0 \\ 0 & -\one
\end{pmatrix}
\;.
$$
 Let us begin by outlining how the skew localizer is constructed for all these cases:

\begin{enumerate}

\item Build the even or odd spectral localizer $L_\kappa$ for the pairing just as in Section~\ref{sec-complexpairing}. For pairings of a projection with a unitary, there is no freedom, but for pairings of two projections or two unitaries one can take either the even or the odd spectral localizer (and it is shown below in Proposition~\ref{prop-Pairings2P2U} that they are conjugate to each other). For pairings of two projections $P$ and $E$, let us work with the even spectral localizer given by \eqref{eq-evenLoc} with $P=\chi(H\leq 0)$ and $E=\chi(D_0=D_0^*\geq 0)$. For pairings of two unitaries $U=A|A|^{-1}$ and $F$, let us consider $T'$ given by \eqref{eq-UniUni}. \Red{Then $D_0$ satisfies $D_0=-\sigma_3 D_0\sigma_3$, so that  $D_0|D_0|^{-1}=\binom{0\;\;F}{F^*\;0}$ and  $A$ is doubled to $A\otimes \sigma_0$, namely $(A\otimes \sigma_0)|A\otimes \sigma_0|^{-1}=\binom{U\;\;0}{0\;\;U}$.} Then the odd spectral localizer is of the \Red{form}
\begin{equation}
\label{eq-OddLocUnitaries}
L^\odd_\kappa
\;=\;
\begin{pmatrix}
\kappa D_0 & A\otimes \sigma_0 \\
A^*\otimes \sigma_0 & -\kappa D_0
\end{pmatrix}
\;.
\end{equation}

\item \Red{Determine an even symmetry operator $Q$ such that
\begin{equation}
\label{eq-ImaginaryLoc}
Q^*\,\overline{L_\kappa}\,Q
\;=\;-L_\kappa
\;.
\end{equation}
For pairings of a projection and a unitary and for pairings of two projection the grading of $\Hh\oplus\Hh$ is chosen such that the spectral localizer is of the form \eqref{eq-evenLoc} and \eqref{eq-oddLoc} given in Section~\ref{sec-complexpairing}. In this grading, $Q$ is given by a $2\times 2$ matrix with entries given by $\SymProd$, $-\SymProd$ or $0$, where 
$$
\SymProd\;=\;\Sigma\,S
\;
$$
is the product of the symmetry operators specifying the case $(j,d)$.  For pairings of two unitaries, namely the cases $(j,d)\in\{(1,0),(5,4),(3,2),(7,6)\}$, the grading of $\Hh\oplus\Hh$ is chosen such that the spectral localizer is of the form \eqref{eq-OddLocUnitaries}. In this grading, $Q$ is given by a $4\times 4$ matrix with entries given by $\SymProd$, $-\SymProd$ or $0$. Further down, explicit formulas for $Q$ are given in all cases.}
  
\item Determine a unitary root $R$ of $Q$ satisfying
\begin{equation}
\label{eq-UnitRoot}
R^2\;=\;Q\;,
\qquad
\overline{R}\;=\;R^*
\;.
\end{equation}

\item Then by construction, \eqref{eq-ImaginaryLoc} and \eqref{eq-UnitRoot} imply that $R^*L_\kappa R$ is purely imaginary so that the skew localizer defined by 
\begin{equation}
\label{eq-SkewLocDef}
\widehat{L}_\kappa
\;=\;
\imath\, R^*L_\kappa R 
\end{equation}
in real and skew-adjoint.

\item In the cases where $j+d$ is odd (namely for pairings of two projections or pairings of two unitaries), Proposition~\ref{prop-OffDiagSkewLoc}  shows that one can construct a further real unitary basis change $N$ such that $N^*\widehat{L}_\kappa N$ is off-diagonal (as for any operator with chiral symmetry).

\end{enumerate}

\begin{proposi}
\label{prop-Pairings2P2U}
For pairings of two projections or pairings of two unitaries, there exists a unitary basis change $M$ such that $L^\odd_{\kappa}=M^*L^\even_{\kappa}M$ and, moreover,  the spectral localizers are chiral, that is, there exists a symmetry $J$ such that $J^*{L}_\kappa J=-{L}_\kappa$. 
\end{proposi}

\noindent {\bf Proof.} For pairings of two projections $P$ and $E$, the even spectral localizer is given by \eqref{eq-evenLoc} with $P=\chi(H\leq 0)$ and $E=\chi(D_0=D_0^*\geq0)$. One has $L^\odd_{\kappa}=M^*L^\even_{\kappa}M$ for $M$ as in \eqref{eq-PassageOddEven} and $\sigma_2^*L^\even_\kappa \sigma_2=-L^\even_\kappa$.  For pairings of two unitaries $U=A|A|^{-1}$ and $F$, the odd spectral localizer \Red{is} as described in Step 1. Then the even spectral localizer and the basis change are 
\begin{equation}
\label{eq-EvenLocUnitaries}
L^\even_{\kappa}
\;=\;
\begin{pmatrix}
0 & -A & \kappa D_1^* & 0 \\ 
-A^* & 0 & 0 & \kappa D_1^* \\ 
\kappa D_1 & 0 & 0 & A \\ 
0 & \kappa D_1 & A^* & 0 
\end{pmatrix}\,,
\qquad
M\;=\;
\begin{pmatrix}
0 & \one & 0 & 0 \\ 
0 & 0 & 0 & -\one \\ 
\one & 0 & 0 & 0 \\ 
0 & 0 & \one & 0 
\end{pmatrix}\,,
\end{equation}
where $D_0=\binom{0\;\;D_1}{D_1^*\;0}$, namely one directly checks that $L^\odd_{\kappa}=M^*L^\even_{\kappa}M$.
As to the chirality, one can verify that $\sigma_3\otimes \sigma_3 L^\odd_\kappa \sigma_3\otimes \sigma_3=-L^\odd_\kappa$.
\hfill $\Box$

\begin{proposi}
\label{prop-OffDiagSkewLoc}
For pairings of two projections or pairings of two unitaries, the skew localizer is chiral, that is, there exists a symmetry $I$ such that $I^*\widehat{L}_\kappa I=-\widehat{L}_\kappa$. 
\end{proposi}

\noindent {\bf Proof.} The argument requires the knowledge of the basis change $R$ in Step 3 which enters into the definition of the skew localizer. It is given case by case in the following subsections (independent of Proposition~\ref{prop-OffDiagSkewLoc}), and is used here. 
For pairings of two unitaries dealt with in Sections~\ref{sec-jd10} and \ref{sec-jd32}, the symmetry $I$ of Proposition~\ref{prop-Pairings2P2U} commutes with $R$ in Step 3 of the construction of the skew localizer. Therefore the claim with $I=J$ is a direct consequence of Proposition~\ref{prop-Pairings2P2U}. Pairings of two projections are dealt with in Sections~\ref{sec-jd07} and \ref{sec-jd21} which now also have to be treated separately. For $(j,d)\in\{(2,1),(6,5)\}$, let us focus on the odd spectral localizer. For $R$ given by \eqref{eq-QRj0d7}, $R^*\sigma_2R=\sigma_3$. Therefore
$$
\sigma_3^*R^*L_\kappa^\odd R \sigma_3
\;=\;
R^*\sigma_2RR^*L_\kappa^\odd RR^*\sigma_2R
\;=\;
R^*\sigma_2L_\kappa^\odd\sigma_2R
\;=\;
-R^*L_\kappa^\odd R\;.
$$
Therefore $\widehat{L}_\kappa$ is chiral w.r.t. $I=\sigma_3$. For $(j,d)\in\{(0,7),(4,3)\}$ the argument is similar.
\hfill $\Box$

\vspace{.2cm}

Steps 1~to~5 being accomplished, one next has to construct finite volume restrictions of the skew localizer $\widehat{L}_{\kappa,\rho}$. While this is similar as for the spectral localizer in Section~\ref{sec-complexpairing}, one rather uses the spectral decomposition of $\widehat{D}=\imath R^*D R $ and sets
$$
(\Hh\oplus\Hh)^\wedge_\rho\;=\;\Ran(\chi(|\widehat{D}|\leq \rho))
\;,
$$
which is finite dimensional because $\widehat{D}$ has compact resolvent.
As $\chi(|\widehat{D}|\leq \rho)=R^*\chi(|D|\leq\rho)R$ and $\overline{|D|}=Q^*|D|Q$, one has
\begin{align*}
\overline{\chi(|\widehat{D}|\leq \rho)}
\;=\;
R\,\overline{\chi(|D|\leq\rho)}\,R^*\;=\;R\,Q^*\,\chi(|D|\leq\rho)\,Q\,R^*
\;=\;R^*\,\chi(|D|\leq\rho)\,R
\;=\;\chi(|\widehat{D}|\leq \rho)
\;.
\end{align*}
Therefore $(\Hh\oplus\Hh)^\wedge_\rho$ is invariant under complex conjugation. If now $\widehat{\pi}_\rho:\Hh\oplus\Hh\to (\Hh\oplus\Hh)^\wedge_\rho$ denotes the associated partial isometry, the finite volume restriction of the skew localizer is
$$
\widehat{L}_{\kappa,\rho}\;=\;\widehat{\pi}_\rho\, \widehat{L}_\kappa \,(\widehat{\pi}_\rho)^*
\;,
$$
which is a real skew-adjoint matrix acting on $(\Hh\oplus\Hh)^\wedge_\rho$. Similarly also $\widehat{D}_{\rho}=\widehat{\pi}_\rho \widehat{D} (\widehat{\pi}_\rho)^*$ is real and skew-adjoint. Let us note that there is no natural decomposition of $\widehat{\pi}_\rho$ into a sum of two partial isometries (as for $\pi_\rho=\pi_\rho\oplus\pi_\rho$  in Section~\ref{sec-complexpairing}). Nevertheless, there is a relation linking $\pi_\rho$ and $\widehat{\pi}_\rho$, namely $R^*\pi_\rho=\widehat{\pi}_\rho R^*$. This implies that the finite volume skew localizer can be expressed in terms of the finite volume spectral localizer via
$$
\widehat{L}_{\kappa,\rho}\;=\;\imath\,R^*_\rho \,L_{\kappa,\rho}\, R_\rho
\;,
\qquad
R_\rho\;=\;\pi_\rho \,R\,(\widehat{\pi}_\rho)^*
\;,
$$
so that it is merely built from the Dirichlet restrictions of $H$. Furthermore, in the situation of Proposition~\ref{prop-OffDiagSkewLoc}, the symmetry $I$ satisfies $I(\widehat{\pi}_\rho)^*=(\widehat{\pi}_\rho)^*\widehat{\pi}_\rho I(\widehat{\pi}_\rho)^*$ because $I\widehat{D}I=-\widehat{D}$. Thus with $I_\rho=\widehat{\pi}_\rho I(\widehat{\pi}_\rho)^*$, one has $I_\rho\widehat{L}_{\kappa,\rho}I_\rho=-\widehat{L}_{\kappa,\rho}$, that is, the finite volume skew localizer is chiral w.r.t. $I_\rho$ just as $\widehat{D}_\rho$. Hence there is a basis change $N_\rho$ such that
\begin{equation}
\label{eq-SkewLocOffdiag}
N_\rho^*\widehat{L}_{\kappa,\rho}N_\rho
\;=\;
\begin{pmatrix}
0 & -B_\rho^* \\ B_\rho & 0
\end{pmatrix}
\;,
\qquad
N_\rho^*\widehat{D}_\rho N_\rho
\;=\;
\begin{pmatrix}
0 & -C_\rho^* \\ C_\rho & 0
\end{pmatrix}
\;.
\end{equation}

It will always be assumed without further mention that $\kappa$ and $\rho$ satisfy either the bounds \eqref{eq-rhoevenSpecLocd0j2} or the bounds \eqref{eq-kappa0odd},  pending on whether the even or odd spectral localizer is used. Because the skew localizer is merely a unitary basis change of the spectral localizer, this implies that it is invertible so that its Pfaffian does not vanish. In each of the $16$ cases with $\ZM_2$-index in Theorem~\ref{theo-indexlist}, the aim is then to prove a formula linking the corresponding $\ZM_2$-index $\Ind_2(T)$ to the sign of the Pfaffian of the finite volume skew localizer: 
\begin{equation}
\label{eq-SkewLocPfaf}
\Ind_2(T)
\;=\;
\sgn\big(\Pf(\widehat{L}_{\kappa,\rho})\big)
\,
\sgn\big(\Pf(\widehat{D}_{\rho})\big)
\;.
\end{equation}
If $j+d$ is odd, then plugging \eqref{eq-SkewLocOffdiag} into \eqref{eq-SkewLocPfaf} one obtains 
\begin{equation}
\label{eq-SkewLocDet}
\Ind_2(T)
\;=\;
\sgn\big(\det(B_\rho)\big)
\,
\sgn\big(\det(C_\rho))\big)
\;.
\end{equation}
Explicit expressions for $B_\rho$ and $C_\rho$ will be given below. The final outcome of the analysis for the special situation $S=\Sigma=\one$ of the case $(j,d)=(2,1)$ is given by formulas \eqref{eq-IntroMainResPf} and \eqref{eq-IntroMainRes} stated in the introduction. Some of the $16$ $\ZM_2$-entries of the table in Theorem~\ref{theo-indexlist} can be treated simultaneously and are thus regrouped in the following subsections.

\subsection{Skew localizer for $(j,d)\in\{(1,7),(5,3)\}$}
\label{sec-j1d7}

For $(j,d)\in\{(1,7),(5,3)\}$, the index pairing is $T=EUE+\one-E$ with $E$ being even Lagrangian and $U$  even real w.r.t. the even symmetry operator $\SymProd=\Sigma S$.  Correspondingly, the odd spectral localizer $L^\odd_\kappa$ has to be used with $D_0=D_0^*=-\Sigma^* \overline{D_0}\Sigma=- \SymProd^* \overline{D_0}\SymProd$. It satisfies \eqref{eq-ImaginaryLoc} with the even symmetry operator $Q$ given by
\begin{equation}
\label{eq-Q-j1d7}
Q\;=\;
\begin{pmatrix}
\SymProd & 0 \\ 0 & -\SymProd
\end{pmatrix}
\;.
\end{equation}
Its root $R$ satisfying \eqref{eq-UnitRoot} is given by
\begin{equation}
\label{eq-R-j1d7}
R
\;=\;
\begin{pmatrix}
\Rr & 0 \\ 0 & \imath\,\Rr
\end{pmatrix}
\;,
\end{equation}
where here and below
$$
\Rr^2\;=\;\Ss
\;,
\qquad
\overline{\Rr}\;=\;\Rr^*
\;.
$$
Note that $\Rr$ can be written as a product of the roots of $S$ and $\Sigma$. Following the strategy outlined above, the skew localizer 
\begin{equation}
\label{eq-SkewLoc-d3j5}
\widehat{L}_\kappa
\;=\;
\imath\, R^*L^\odd_\kappa R
\;=\;
\begin{pmatrix}
\imath\,\kappa\,\Rr^*\,D_0\,\Rr & -\,\Rr^*\,A\,\Rr  \\
\Rr^*\,A^*\,\Rr & -\,\imath\,\kappa\,\Rr^*\,D_0\,\Rr
\end{pmatrix}
\;,
\end{equation}
is real and skewadjoint. The same holds for $\widehat{D}=\imath R^*D^\odd R $.

\begin{theo}
\label{theo-SpecLocd3j5}
For $(j,d)\in\{(1,7),(5,3)\}$, the identity \eqref{eq-SkewLocPfaf} holds with $\widehat{L}_\kappa$ given in \eqref{eq-SkewLoc-d3j5}.
\end{theo}

\vspace{.2cm}

\noindent {\bf Proof.}
As in prior works \cite{LS1,LS2,LS3,LSS}, it is necessary to show that the r.h.s. of \eqref{eq-SkewLocPfaf}  is independent of $\kappa$ and $\rho$, as long as they satisfy the bounds \eqref{eq-kappa0odd}. For a variation of $\kappa$, this immediately follows from the invertibility of $L^\odd_{\kappa,\rho}$ (and thus $\widehat{L}_{\kappa,\rho}$) guaranteed by Theorem~\ref{tho-LS2}. Changing $\rho$ leads to jumps of the dimension of the finite volume Hilbert spaces $(\Hh\oplus\Hh)^\wedge_\rho$ and this requires a more careful analysis. This can be carried out by following very closely the argument of Section~2 of \cite{LS3}. In fact, the skew localizer is obtained from the spectral localizer by conjugation with $R$ which is compatible with the finite volume restrictions. In the last part of the argument, one then can replace the signature of the spectral localizer by the sign of the Pfaffian of the skew localizer.

\vspace{.2cm}

Let us now turn to the proof of the identity \eqref{eq-SkewLocPfaf} for which $\kappa>0$ can be chosen as small as needed and $\rho$ as large as needed. Let us consider the odd increasing differentiable function $F_1: \RM \to \RM$ given by
$$
F_1(x)\;=\;
\begin{cases}
-2\;, & x<-2\;, \\
-x^3-4x^2-4x-2\;, & x\in[-2,-1]\;,\\
x\;, & x\in[-1,1]\;,\\
-x^3+4x^2-4x+2\;, & x\in[1,2]\;,\\
2\;, & x>2\;.
\end{cases}
$$
The Fourier transform $\widehat{F'_1}$ of the derivative $F'_1$ can be computed explicitly to be
$$
\widehat{F'_1}(p)\;=\;\tfrac{1}{\pi}\big(\tfrac{-4\cos(2p)}{p^2}+\tfrac{-2\cos(p)}{p^2}+\tfrac{6\sin(2p)}{p^3}+\tfrac{-6\sin(p)}{p^3}\big)
\;.
$$ 
Hence one has an $L^1$-norm bound $\|\widehat{F'_1}\|_1\leq \tfrac{28}{\pi}$. Let us scale to $F_\rho: \RM \to \RM$ given by
\begin{equation}
\label{eq-defFroh}
F_\rho(x)
\;=\;
\rho\, F_1(\tfrac{x}{\rho})
\;.
\end{equation} 
Then $F_\rho$ is an odd increasing differentiable function with $F_\rho(x)=x$ for $|x|\leq \rho$ and $F_\rho(x)=2\rho=-F_\rho(-x)$ for $|x|\geq2\rho$. Furthermore the $L^1$-norm of the Fourier transform of the derivative is still bounded by $\tfrac{28}{\pi}$. As in Proposition~4 in \cite{LS1}, standard bounds imply
\begin{equation}
\label{eq-boundFroh}
\|[F_\rho(D_0),A]\|\;\leq\;\frac{28}{\pi} \|[D_0,A]\|\;,\qquad 
\|[F_\rho(D_0),U]\|\;\leq\;\frac{28}{\pi} \|[D_0,U]\|
\;.
\end{equation}
Moreover, $F_\rho(D_0)$ inherits the symmetry of $D_0$
$$
\Ss^*\,\overline{F_\rho(D_0)}\,\Ss\;=\;-F_\rho(D_0)\;.
$$
As $[F_\rho(D),U]$ is compact, the linear path connecting $\imath F_\rho(D_0)$ to $U\imath F_\rho(D_0)U\Red{^*}$ lies in $\thetaSFeven(E,U,\Ss)$. By Corollary~\ref{coro-index=flowd3j5} and the homotopy invariance of the orientation flow, the index $\Ind_2=\Ind_2(EUE+\one-E)$ is given by
\begin{align*}
\Ind_2
\;=\;
\Of_2(t\in[0,1]\mapsto (1-t)\imath F_\rho(D_0)+tU\imath F_\rho(D_0)U^*)\;.
\end{align*}
In the following we denote the orientation flow along the straight line path connecting two operators $T_0$ and $T_1$  by $\Of_2(T_0, T_1)$ whenever it is well-defined. Because the second summand is trivial, the additivity of the orientation flow now leads to
\begin{align*}
\Ind_2
&
\;=\;\Of_2\left(\imath 
\begin{pmatrix}
\kappa F_\rho(D_0) & 0\\
0 & -\kappa F_\rho(D_0) 
\end{pmatrix},\imath 
\begin{pmatrix}
U & 0\\
0 & \one
\end{pmatrix}
\begin{pmatrix}
\kappa F_\rho(D_0) & 0\\
0 & -\kappa F_\rho(D_0) 
\end{pmatrix}
\begin{pmatrix}
U & 0\\
0 & \one
\end{pmatrix}^*
\right)
\\
&
\;=\;\Of_2\left(\imath 
\begin{pmatrix}
\kappa F_\rho(D_0) & 0\\
0 & -\kappa F_\rho(D_0) 
\end{pmatrix},\imath 
\begin{pmatrix}
U & 0\\
0 & \one
\end{pmatrix}
\begin{pmatrix}
\kappa F_\rho(D_0) & \one\\
\one & -\kappa F_\rho(D_0) 
\end{pmatrix}
\begin{pmatrix}
U & 0\\
0 & \one
\end{pmatrix}^*
\right)
\;,
\end{align*}
where the second step follows from the homotopy invariance of the $\ZM_2$-valued spectral flow because
$$
s\in[0,1]\mapsto \imath R^*
\begin{pmatrix}
\kappa F_\rho(D_0) & s\one\\
s\one & -\kappa F_\rho(D_0) 
\end{pmatrix}R
$$
is a continuous path of real skew-adjoint invertibles and the linear path connecting
$$
\begin{pmatrix}
\kappa F_\rho(D_0) & 0\\
0 & -\kappa F_\rho(D_0) 
\end{pmatrix}
\qquad
\mbox{to}
\qquad
\begin{pmatrix}
U & 0\\
0 & \one
\end{pmatrix}
\begin{pmatrix}
\kappa F_\rho(D_0) & s\one\\
s\one & -\kappa F_\rho(D_0) 
\end{pmatrix}
\begin{pmatrix}
U & 0\\
0 & \one
\end{pmatrix}^*
$$
is within the Fredholm operators for all $s \in [0,1]$ as $[F_\rho(D),U]$ is compact. Multiplying out shows
\begin{align*}
\Ind_2
\;=\;
\Of_2\left(\imath 
\begin{pmatrix}
\kappa F_\rho(D_0) & 0\\
0 & -\kappa F_\rho(D_0) 
\end{pmatrix},\imath 
\begin{pmatrix}
\kappa U F_\rho(D_0) U^* & U\\
U^* & -\kappa F_\rho(D_0) 
\end{pmatrix}
\right)\;.
\end{align*}
For $\kappa$ sufficiently small, the linear path from
$$
\begin{pmatrix}
\kappa U F_\rho(D_0) U^* & U\\
U^* & -\kappa F_\rho(D_0) 
\end{pmatrix}
\qquad
\mbox{to}
\qquad
\begin{pmatrix}
\kappa  F_\rho(D_0)  & U\\
U^* & -\kappa F_\rho(D_0) 
\end{pmatrix}
$$
is within the invertibles because of the bound \eqref{eq-boundFroh}. As $[F_\rho(D_0), U]$ is compact, the homotopy invariance of the orientation flow implies
\begin{align*}
\Ind_2
\;=\;
\Of_2\left(\imath 
\begin{pmatrix}
\kappa F_\rho(D_0) & 0\\
0 & -\kappa F_\rho(D_0) 
\end{pmatrix},\imath 
\begin{pmatrix}
\kappa F_\rho(D_0)  & U\\
U^* & -\kappa F_\rho(D_0) 
\end{pmatrix}
\right)\;.
\end{align*}
For $\kappa$ sufficiently small,
$$
s\in[0,1]\mapsto \imath R^*
\begin{pmatrix}
\kappa F_\rho(D_0) & U|A|^s\\
(U|A|^s)^* & -\kappa F_\rho(D_0) 
\end{pmatrix}R
$$
is a continuous path of real skew-adjoint invertibles. As $[F_\rho(D_0),  U|A|^s]$ is compact for all $s \in [0,1]$,
\begin{align*}
\Ind_2
\;=\;
\Of_2\left(\imath \kappa F_\rho(D),\imath \widetilde{L}^\odd_\kappa \right)\;,
\end{align*}
where
$$
\widetilde{L}^\odd_\kappa
\;=\;
\begin{pmatrix}
\kappa F_\rho(D_0) & A\\
A^* & -\kappa F_\rho(D_0) 
\end{pmatrix}
\;.
$$
For $\Hh_{\rho^c}=\Hh\ominus \Hh_\rho$ we denote the surjective partial isometry onto $\Hh_{\rho^c}$ and $\Hh_{\rho^c}\oplus \Hh_{\rho^c}$  by $\pi_{\rho^c}$. For any operator $B$ on $\Hh$ or $\Hh \oplus \Hh$  let us define $B_{\rho^c}=\pi_{\rho^c}B(\pi_{\rho^c})^*$. One has $F_\rho(D_0)=F_\rho(D_0)_\rho \oplus F_\rho(D_0)_{\rho^c}$ and $F_\rho(D_0)_\rho=(D_0)_\rho$ and similarly for $D$. Thus also $\widetilde{L}^\odd_{\kappa,\rho}={L}^\odd_{\kappa,\rho}$.  Next we show that the linear path
$$
t\in [0,1]\; \mapsto\; \widetilde{L}^\odd_\kappa(t)\;=\; \begin{pmatrix}
L^\odd_{ \kappa,\rho} & 0\\
0 & \widetilde{L}^\odd_{ \kappa,\rho^c}
\end{pmatrix}\;+\;t\,
\begin{pmatrix}
0 & \pi_\rho H (\pi_{\rho^c})^*\\
\pi_{\rho^c} H (\pi_{\rho})^* & 0 
\end{pmatrix}
$$
is within the invertibles. Let us first check that 
$$
\widetilde{L}^\odd_{\kappa,\rho^c}\;=\;\begin{pmatrix}
\kappa F_\rho(D_0)_{ \rho^c}  & A_{ \rho^c}\\
A^*_{ \rho^c} & -\kappa F_\rho(D_0)_{ \rho^c} 
\end{pmatrix}
$$
is invertible. One has
\begin{align*}
(\widetilde{L}^\odd_{\kappa,\rho^c})^2\;&=\;
\begin{pmatrix}
\kappa^2 F_\rho(D_0)_{ \rho^c}^2 +A_{ \rho^c}(A_{ \rho^c})^* & \kappa(F_\rho(D_0)A_{ \rho^c}-A_{ \rho^c}F_\rho(D_0))\\
\kappa(F_\rho(D_0)A_{ \rho^c}-A_{ \rho^c}F_\rho(D_0))^* & \kappa^2 F_\rho(D_0)_{ \rho^c}^2+ (A_{ \rho^c})^*A_{ \rho^c}
\end{pmatrix}\\
\;&\geq\; (\kappa^2\rho^2-\kappa \|[F_\rho(D_0),A]\|)\one\\
\;&\geq\;(\kappa^2\rho^2-\kappa\tfrac{28}{\pi} \|[D_0,A]\|)\one\\
\;&\geq\tfrac{1}{2}\kappa^2\rho^2\one
 \end{align*}
where the third step follows from \eqref{eq-boundFroh} and the last one from \eqref{eq-kappa0odd}. Hence $L^\odd_\kappa(t)$ is given by
$$
|L^\odd_{\kappa, \rho}\oplus \widetilde{L}^\odd_{\kappa, \rho^c}|^\frac{1}{2}\left(G+t
 \begin{pmatrix}
0 & \!\!\!\!\!\! |L^\odd_{\kappa, \rho}|^{-\frac{1}{2}}\pi_\rho H(\pi_{\rho^c})^* |\widetilde{L}^\odd_{\kappa, \rho^c}|^{-\frac{1}{2}}\\
|\widetilde{L}^\odd_{\kappa, \rho^c}|^{-\frac{1}{2}}\pi_{\rho^c} H(\pi_\rho)^* |L^\odd_{\kappa, \rho}|^{-\frac{1}{2}} \!\!\!\!\!\! & 0
\end{pmatrix}\right) |L^\odd_{\kappa, \rho}\oplus \widetilde{L}^\odd_{\kappa, \rho^c}|^\frac{1}{2}
$$
where $G$ is a diagonal unitary w.r.t. the direct sum $\Hh=\Hh_\rho\oplus \Hh_{\rho^c}$. The off-diagonal entries satisfy
$$
\left\||L^\odd_{\kappa, \rho}|^{-\frac{1}{2}}\pi_\rho H(\pi_{\rho^c})^* |\widetilde{L}^\odd_{\kappa, \rho^c}|^{-\frac{1}{2}}\right\|\leq\frac{\sqrt[4]{8}\|H\|}{\sqrt{\kappa\rho g}}\;,
$$
thus they are smaller than $1$ for $\rho$ sufficiently large.
As $L^\odd_{ \kappa}-(L^\odd_{ \kappa,\rho}\oplus \widetilde{L}^\odd_{ \kappa,\rho^c})$
 is compact, the homotopy invariance of the orientation flow implies
\begin{align*}
\Ind_2
&\;=\;
\Of_2\left(\imath \kappa F_\rho(D),
 \imath (L^\odd_{\kappa, \rho}\oplus L^\odd_{\kappa, \rho^c})\right)
 \;=\;
\Of_2\left(\imath \kappa F_\rho(D)_\rho, L^\odd_{\kappa, \rho} \right)
\;+ \;\Of_2\left(\imath \kappa F_\rho(D)_{\rho^c}, L^\odd_{\kappa, \rho^c} \right)\;.
\end{align*}
Now $(\kappa F_\rho(D)_{\rho^c})^2\geq \kappa^2\rho^2$ so that the path $t \in [0,1]\mapsto t  \kappa F_\rho(D)_{\rho^c} + (1-t)\widetilde{L}^\odd_{\kappa, \rho^c}$ consists of invertibles for $\rho$ sufficiently large. Therefore using $ F_\rho(D) _\rho=D_\rho$
\begin{align*}
\Ind_2
&\;=\;
\Of_2\left(\imath \kappa F_\rho(D)_\rho, L^\odd_{\kappa, \rho} \right)
\;=\;
\SF_2\left(\kappa \widehat{D}_\rho, \widehat{L}_{\kappa,\rho}
\right)
\;=\;\sgn\big(\Pf(\widehat{L}_{\kappa,\rho})\big)
\,
\sgn\big(\Pf(\widehat{D}_{\rho})\big)\;,
\end{align*}
completing the proof.
\hfill$\Box$

\subsection{Skew localizer for $(j,d)\in\{(3,1),(7,5)\}$}
\label{sec-j3d1}

As in Section~\ref{sec-j1d7},  the index pairing is of the form $T=EUE+\one-E$, but the symmetry operator $\Ss=\Sigma S$ is odd and $\one-2E$ and $U$ are odd symmetric w.r.t. $\Ss$ which after a suitable basis change is of the form $\Ss=\imath \sigma_2$. The even symmetry operator $Q$ and its root $R$ are
\begin{equation}
\label{eq-Q-j3d1}
Q\;=\;\begin{pmatrix}
0 & \Ss 
\\
-\Ss & 0
\end{pmatrix}
\;,
\qquad
R
\;=\;
\frac{1+\imath}{2}
\begin{pmatrix}
\sigma_0 & \sigma_2 \\ -\sigma_2 & \sigma_0
\end{pmatrix}
\;.
\end{equation}
Hence one finds for the skew localizer
\begin{equation}
\label{eq-LXYZ}
\widehat{L}_\kappa
\;=\;
\begin{pmatrix}
X & -Y\Red{^T} \\ Y & Z
\end{pmatrix}
\;,
\end{equation}
where $X=\frac{\imath}{2}(\sigma_0 \kappa D_0\sigma_0-\sigma_2 \kappa D_0\sigma_2-\sigma_0 A\sigma_2-\sigma_2 A^*\sigma_0)$, $Y=\frac{\imath}{2}(\sigma_2 \kappa D_0\sigma_0+\sigma_0 \kappa D_0\sigma_2-\sigma_2 A\sigma_2+\sigma_0 A^*\sigma_0)$ and  $Z=\frac{\imath}{2}(-\sigma_0 \kappa D_0\sigma_0+\sigma_2 \kappa D_0\sigma_2+\sigma_0 A^*\sigma_2+\sigma_2 A\sigma_0)$. 

\begin{theo}
\label{theo-speclocd1j3}
For $(j,d)\in\{(3,1),(7,5)\}$, the identity \eqref{eq-SkewLocPfaf} holds with $\widehat{L}_\kappa$ given in \eqref{eq-LXYZ}.
\end{theo}

\noindent {\bf Proof.}
 The argument showing that the r.h.s. of \eqref{eq-SkewLocPfaf} is independent of $\kappa$ and $\rho$ if the bounds \eqref{eq-kappa0odd} hold is the same as in the proof of Theorem~\ref{theo-SpecLocd3j5} and is therefore omitted. It remains to show \eqref{eq-SkewLocPfaf} for some $\kappa$ and $\rho$ such that the bounds \eqref{eq-kappa0odd} hold.  Let us consider the odd increasing differentiable function $F_\rho: \RM \to \RM$, defined by \eqref{eq-defFroh}. Then $\|[F_\rho(D_0),A]\|$ and $\|[F_\rho(D_0),U]\|$ satisfy the bound \eqref{eq-boundFroh}.  Moreover $F_\rho(D_0)$ inherits the symmetry of $D_0$, namely $F_\rho(D_0)$ is odd real w.r.t. $\Ss$.
By Corollary~\ref{coro-index=flowd1j3} $\Ind_2=\Ind_2(EUE+\one-E)$ is given by
$$
\Ind_2
\;=\;
\HSF(E,U)\;=\;\SF(t \in [0,\tfrac{1}{2}]\mapsto (1-t)F_\rho(D_0)+tUF_\rho(D_0)U\Red{^*})\;\mbox{\rm mod}\;2\;.
$$
If $\tfrac{1}{2}(F_\rho(D_0)+UF_\rho(D_0)U\Red{^*})$ is not invertible, $\tfrac{1}{2}(F_\rho(D_0)+UF_\rho(D_0)U\Red{^*})+\epsilon\one$ is invertible for $\epsilon>0$ sufficiently small. Replacing $F_\rho(D_0)$ by $F_\rho(D_0)+\epsilon\one$ does not change the r.h.s. of \eqref{eq-SkewLocPfaf} for $\rho$ sufficiently large and $\epsilon$ sufficiently small. Thus by identifying $F_\rho(D_0)$ with $F_\rho(D_0)+\epsilon\one$ without loss of generality one my assume that $\tfrac{1}{2}(F_\rho(D_0)+UF_\rho(D_0)U\Red{^*})$ is invertible. By Proposition \ref{prop-oddhlaf}, the half-spectral flow $\HSF(E,U)$ is equal to
\begin{align*}
\Of_2\left(\imath 
\begin{pmatrix}
\kappa F_\rho(D_0) & 0\\
0 & -\kappa F_\rho(D_0) 
\end{pmatrix},\imath 
\begin{pmatrix}
\tfrac{\kappa}{2}(F_\rho(D_0)+UF_\rho(D_0)U^*) \!\!\!\!\!\! & 0\\
0 & \!\!\!\!\!\! \tfrac{-\kappa}{2}( F_\rho(D_0)+U^*F_\rho(D_0)U) 
\end{pmatrix}
\right)
\;.
\end{align*}
Therefore
\begin{align*}
&
\Ind_2
\;=\;\Of_2\left(\imath \kappa F_\rho(D)
,\imath 
\begin{pmatrix}
\tfrac{\kappa}{2}(F_\rho(D_0)+UF_\rho(D_0)U^*) \!\!\!\!\!\! & 0\\
0 & \!\!\!\!\!\! \tfrac{-\kappa}{2}( F_\rho(D_0)+U^*F_\rho(D_0)U) 
\end{pmatrix}
\right)\\
&\;=\,
\Of_2\left(\imath 
\kappa F_\rho(D)
,\imath 
\begin{pmatrix}
U & 0\\
0 & \one
\end{pmatrix}
\begin{pmatrix}
\tfrac{\kappa}{2}(F_\rho(D_0)+U^*F_\rho(D_0)U) \!\!\!\!\!\! & 0\\
0 & \!\!\!\!\!\! \tfrac{-\kappa}{2}( F_\rho(D_0)+U^*F_\rho(D_0)U) 
\end{pmatrix}
\begin{pmatrix}
U^* & 0\\
0 & \one
\end{pmatrix}
\right)
\\
&\,=\,
\Of_2\left(\imath 
\kappa F_\rho(D)
,\imath 
\begin{pmatrix}
U & 0\\
0 & \one
\end{pmatrix}
\begin{pmatrix}
\tfrac{\kappa}{2}(F_\rho(D_0)+U^*F_\rho(D_0)U) \!\!\!\!\!\!
& \one\\
\one & \!\!\!\! \!\!\tfrac{-\kappa}{2}( F_\rho(D_0)+U^*F_\rho(D_0)U) 
\end{pmatrix}
\begin{pmatrix}
U^* & 0\\
0 & \one
\end{pmatrix}
\right)
.
\end{align*}
The last step follows from the same homotopy argument as in the proof of Theorem~\ref{theo-SpecLocd3j5}.
Multiplying out leads to 
\begin{align*}
\Ind_2
&\;=\;
\Of_2\left(\imath 
\kappa F_\rho(D),\imath 
\begin{pmatrix}
\tfrac{\kappa}{2}(F_\rho(D_0)+UF_\rho(D_0)U^*)& U\\
U^* & \tfrac{-\kappa}{2}( F_\rho(D_0)+U^*F_\rho(D_0)U) 
\end{pmatrix}
\right)\;.
\end{align*}
For $\kappa$ sufficiently small, the linear path from
$$
\begin{pmatrix}
\tfrac{\kappa}{2}(F_\rho(D_0)+UF_\rho(D_0)U^*)& U\\
U^* & \tfrac{-\kappa}{2}( F_\rho(D_0)+U^*F_\rho(D_0)U) 
\end{pmatrix}
\quad
\mbox{to}
\quad
\begin{pmatrix}
\kappa  F_\rho(D_0)  & U\\
U^* & -\kappa F_\rho(D_0) 
\end{pmatrix}
$$
is within the invertibles since
\begin{align*}
(&1-t)\begin{pmatrix}
\tfrac{\kappa}{2}(F_\rho(D_0)+UF_\rho(D_0)U^*)& U\\
U^* & \tfrac{-\kappa}{2}( F_\rho(D_0)+U^*F_\rho(D_0)U) 
\end{pmatrix}+
t\begin{pmatrix}
\kappa  F_\rho(D_0)  & U\\
U^* & -\kappa F_\rho(D_0) 
\end{pmatrix}\\
\;&=\;
\begin{pmatrix}
\tfrac{\kappa}{2}(F_\rho(D_0)+UF_\rho(D_0)U^*)+\tfrac{t\kappa}{2}[F_\rho(D_0), U]U^* \!\!\!\!\!\! & U\\
U^* & \!\!\!\!\!\! \tfrac{-\kappa}{2}( F_\rho(D_0)+U^*F_\rho(D_0)U) +\tfrac{t\kappa}{2}U^*[F_\rho(D_0), U]
\end{pmatrix}
\end{align*}
and $\|[F_\rho(D_0), U]\|$ is bounded by \eqref{eq-boundFroh}. As $[F_\rho(D_0), U]$ is compact, the homotopy invariance of the orientation flow implies 
\begin{align*}
\Ind_2
\;=\;
\SF_2\left(\imath 
\begin{pmatrix}
\kappa F_\rho(D_0) & 0\\
0 & -\kappa F_\rho(D_0) 
\end{pmatrix},\imath 
\begin{pmatrix}
\kappa F_\rho(D_0)  & U\\
U^* & -\kappa F_\rho(D_0) 
\end{pmatrix}
\right)\;.
\end{align*}
Form now on the argument is again the same as the one leading to Theorem~\ref{theo-SpecLocd3j5}, implying
\begin{align*}
\Ind_2
&\;=\;
\Of_2\left(\imath \kappa F_\rho(D)_\rho, L^\odd_{\kappa, \rho} \right)
\;=\;
\SF_2\left(\kappa \widehat{D}_\rho, \widehat{L}_{\kappa,\rho}
\right)
\;=\;\sgn\big(\Pf(\widehat{L}_{\kappa,\rho})\big)
\,
\sgn\big(\Pf(\widehat{D}_{\rho})\big)\;,
\end{align*}
so that the proof is complete.
\hfill$\Box$

\subsection{Skew  localizer for $(j,d)\in\{(2,0),(6,4)\}$}
\label{sec-j2d0}

For $(j,d)\in\{(2,0),(6,4)\}$, the index pairing is $T=PFP+\one-P$ with $P=\chi(H\leq0)$ and $F=D_0|D_0|^{-1}$. Moreover, $\Ss=\Sigma S$ is an even symmetry operator and $P$ is even Lagrangian and $F$ is even real w.r.t. $\Ss$.  The even symmetry $Q$ and its root $R$ are given by \eqref{eq-Q-j1d7} and \eqref{eq-R-j1d7}. Therefore the skew localizer is 
\begin{equation}
\label{eq-SkewLoc-d2j0}
\widehat{L}_\kappa
\;=\;
\imath\, R^*L^\even_\kappa R
\;=\;
\begin{pmatrix}
-\,\imath\,\Rr^*\,H\,\Rr &-\, \kappa\,\Rr^*\,D_0^*\,\Rr  \\
\,\kappa\,\Rr\,D_0\,\Rr & \imath\,\Rr^*\,H\,\Rr
\end{pmatrix}
\;.
\end{equation}

\begin{theo}
\label{theo-SpecLocd0j2}
For $(j,d)\in\{(2,0),(6,4)\}$, the identity \eqref{eq-SkewLocPfaf} holds with $\widehat{L}_\kappa$ given in \eqref{eq-SkewLoc-d2j0}.
\end{theo}

\noindent {\bf Proof.}
As in the proof of Theorem \ref{theo-SpecLocd3j5} we need to show that the the r.h.s. of \eqref{eq-SkewLocPfaf} is independent of $\kappa$ and $\rho$, as long as they satisfy the bounds \eqref{eq-rhoevenSpecLocd0j2}. Again this follows from a modification of a prior work, namely the proof of Theorem 2 in \cite{LS2} in which the signature of the spectral localizer is replaced by the sign of the Pfaffian of the skew localizer. It remains to show \eqref{eq-SkewLocPfaf} for some $\kappa$ and $\rho$ such that the bounds \eqref{eq-rhoevenSpecLocd0j2} hold. By Corollary~\ref{coro-index=Z2flowd0j2} $\Ind_2=\Ind_2(PFP+\one-P)$ is given by
\begin{align*}
\Ind_2
&\;=\;
\Of_2(t\in[0,1]\mapsto (1-t)\imath H+t\imath F HF^*)\;.
\end{align*}
Due to the additivity of the orientation flow,
\begin{align*}
\Ind_2
\;=\;
\Of_2\left(\imath (-H\otimes \Gamma), \imath \begin{pmatrix}
\one & 0\\
0 & F 
\end{pmatrix}(-H\otimes \Gamma)\begin{pmatrix}
\one & 0\\
0 & F 
\end{pmatrix}^*\right)
\;\Red{,}
\end{align*}
\Red{where $\Gamma=\diag(\one,-\one)$.}
Let us consider the odd increasing differentiable function $F_\rho: \RM \to \RM$, defined by \eqref{eq-defFroh}.
As the $L^1$-norm of the Fourier transform of the derivative is bounded by $\tfrac{28}{\pi}$,
\begin{align}
\label{eq-boundFroheven}
\|[F_\rho(D),H\oplus H]\|\;\leq\;\frac{28}{\pi} \|[D,H\oplus H]\|\;.
\end{align}
Moreover, $F_\rho(D)$ anticommutes with $\Gamma$, hence is of the form
\begin{equation}
\label{eq-FDform}
F_\rho(D)
\;=\;
\begin{pmatrix}
0 & (D_0')^*\\
D_0' & 0
\end{pmatrix}
\end{equation}
and it also inherits the symmetry of $D$, namely $Q^*\overline{F_\rho(D)}Q=-F_\rho(D)$. One has
$$
(-H\otimes \Gamma+t\kappa F_\rho(D))^2\;=\;\begin{pmatrix}
H^2+(t\kappa)^2|D_0'|^2 & \Red{-}t\kappa [H,D_0']^*\\
t\kappa [H,D_0'] & H^2+(t\kappa)^2|(D_0')^*|^2 
\end{pmatrix}\geq(g^2-t\kappa\|[F_\rho(D),H\oplus H]\|)\one\;,
$$
for $t \in [0,1]$. By \eqref{eq-boundFroheven}, the straight line path connecting $-H\otimes \Gamma$ to $-H\otimes \Gamma+\kappa F_\rho(D)$ is within the invertibles for $\kappa$ sufficiently small. Hence
$$
\Ind_2
\;=\;
\Of_2\left(\imath (-H\otimes \Gamma+\kappa F_\rho(D)), \imath \begin{pmatrix}
\one & 0\\
0 & F 
\end{pmatrix}(-H\otimes \Gamma)\begin{pmatrix}
\one & 0\\
0 & F 
\end{pmatrix}^*\right)
\;.
$$
Next one directly checks that
\begin{align*}
s\in[0,\kappa\rho]
\;\mapsto\; 
\begin{pmatrix}
\one & 0\\
0 & F 
\end{pmatrix}
\begin{pmatrix}
-H & s\\
s & H
\end{pmatrix}
\begin{pmatrix}
\one & 0\\
0 & F 
\end{pmatrix}^*
\;=\;
\begin{pmatrix}
-H & sF^*\\
sF & FHF^* 
\end{pmatrix}
\end{align*}
is a path of invertibles. Let us also show that 
\begin{align*}
A(s,t)\;&=\;t(-H\otimes \Gamma+\kappa F_\rho(D))
\;+\;
(1-t)\begin{pmatrix}
-H & sF^*\\
sF & FHF^*
\end{pmatrix}\\
\;&=\;
\begin{pmatrix}
-H & t\kappa (D_0')^*+(1-t)sF^*\\
t\kappa D_0'+(1-t)sF & H-(1-t)[H,F]F^*
\end{pmatrix}
\end{align*}
is Fredholm for all $(s,t)\in [0,\kappa\rho]\times [0,1]$. Because $[H,F]$ is compact, it is sufficient to show that 
$$
B(s,t)\;=\;\begin{pmatrix}
-H & t\kappa (D_0')^*+(1-t)sF^*\\
t\kappa D_0'+(1-t)sF & H
\end{pmatrix}
$$
is Fredholm. One can replace $D|D|^{-1}$ by $\tfrac{1}{2\rho}F_\rho(D)$ as $\Ran (\tfrac{1}{2\rho}F_\rho(D)-D|D|^{-1})\subset (\Hh\oplus\Hh)_{2\rho}$ is finite dimensional, so that $\tfrac{1}{2\rho}F_\rho(D)-D|D|^{-1}$ is compact. Therefore it is sufficient to show that
$$
C(s,t)\;=\; -H\otimes \Gamma +t\kappa F_\rho(D)+(1-t)s\tfrac{1}{2\rho}F_\rho(D)
$$
is Fredholm. Now
\begin{align*}
C(s,t)^2\,&=(H\otimes \Gamma)^2+(t\kappa F_\rho(D)+(1-t)s\tfrac{1}{2\rho}F_\rho(D))^2-[t\kappa F_\rho(D)+(1-t)s\tfrac{1}{2\rho}F_\rho(D),H\oplus H]\Gamma\\
&\geq (g^2-(t\kappa+ (1-t)s\tfrac{1}{2\rho})\|[F_\rho(D),H\oplus H]\|)\one\\
&\geq (g^2-(\kappa+\tfrac{\kappa}{2})\|[F_\rho(D),H\oplus H]\|)\one\\
&\geq (g^2-\tfrac{42\kappa}{\pi}\|[D,H\oplus H]\|)\one\;,
\end{align*}
where the last step follows form \eqref{eq-boundFroheven}. Therefore $C(s,t)$ is invertible and $A(s,t)$ is Fredholm for all $(s,t)\in [0,\kappa\rho]\times [0,1]$ and $\kappa$ sufficiently small.
This implies by the homotopy invariance of the orientation flow
\begin{align*}
\Ind_2&
\;=\;
\Of_2\left(\imath \begin{pmatrix}
-H & \kappa\rho F^*\\
\kappa\rho F & FHF^*
\end{pmatrix}, \imath (-H\otimes \Gamma+\kappa F_\rho(D))\right)\;.
\end{align*}
For $(\Hh\oplus\Hh)_{\rho^c}=(\Hh\oplus\Hh)\ominus (\Hh\oplus\Hh)_\rho$ we again denote the surjective partial isometry onto $(\Hh\oplus\Hh)_{\rho^c}$ by $\pi_{\rho^c}$, and for any operator $B$ on $\Hh \oplus \Hh$  set $B_{\rho^c}=\pi_{\rho^c}B(\pi_{\rho^c})^*$. Now the finite volume restriction is done similarly as in the proof of Theorem~\ref{theo-SpecLocd3j5}, with modifications for the even case as in \cite{LS2}. It leads to
\begin{align*}
\Ind_2
\;=\;
\Of_2\left(\imath \begin{pmatrix}
-H & \kappa\rho F^*\\
\kappa\rho F & FHF^*
\end{pmatrix}, \imath ((-H\otimes \Gamma+\kappa F_\rho(D))_{\rho}\oplus (-H\otimes \Gamma+\kappa F_\rho(D))_{\rho^c})\right)\;.
\end{align*}
The path
$$
s\in[0,1]
\;\mapsto\; 
A(s)\;=\;\begin{pmatrix}
-sH & \kappa\rho F^*\\
\kappa\rho F & s\Red{FHF^*}
\end{pmatrix}\\
$$
is within the invertibles for $\rho$ sufficiently large. As $tA(s)_{\rho^c}+(1-t)(-H\otimes \Gamma+\kappa F_\rho(D))_{\rho^c}$ is invertible for all $(s,t)\in[0,1]\times[0,1]$ and $\rho$ sufficiently large,
$$
tA(s)+(1-t)((-H\otimes \Gamma+\kappa F_\rho(D))_{\rho}\oplus (-H\otimes \Gamma+\kappa F_\rho(D))_{\rho^c})
$$
is Fredholm for all $(s,t)\in[0,1]\times[0,1]$, so that
\begin{align*}
\Ind_2
&\;=\;
\Of_2\left(\imath \kappa\rho D|D|^{-1}, \imath ((-H\otimes \Gamma+\kappa F_\rho(D))_{\rho}\oplus (-H\otimes \Gamma+\kappa F_\rho(D))_{ \rho^c})\right)\\
&\;=\;\Of_2\left(\imath \kappa\rho (D|D|^{-1})_\rho , \imath (-H\otimes \Gamma+\kappa F_\rho(D))_{\rho}\right)
\\ &\quad
\;+\;\Of_2\left(\imath  \kappa\rho (D|D|^{-1})_{\rho^c}, \imath (-H\otimes \Gamma+\kappa F_\rho(D))_{ \rho^c})\right)\;.
\end{align*}
The second summand vanishes because the linear path
$$
t\in[0,1]
\;\mapsto\;
\Red{t}\begin{pmatrix}
0 & \kappa\rho F^*\\
\kappa\rho F &0
\end{pmatrix}_{\rho^c}+(1-t)(-H\otimes \Gamma+\kappa F_\rho(D))_{ \rho^c}
$$
lies in the invertibles for $\rho$ sufficiently large. As $(-H\otimes \Gamma+\kappa F_\rho(D))_\rho=L^\even_{\kappa,\rho}$,
\begin{align*}
\Ind_2
&\;=\;
\Of_2\left(\imath \kappa\rho (D|D|^{-1})_\rho , \imath (-H\otimes \Gamma+\kappa F_\rho(D))_{\rho}\right)
\;=\;\sgn\big(\Pf(\widehat{L}_{\kappa,\rho})\big)
\,
\sgn\big(\Pf(\widehat{D}_{\rho})\big)\;,
\end{align*}
completing the proof.
\hfill$\Box$

\vspace{.2cm}

\noindent {\bf Remark:} If $\Of_2$ and $\Ind_2$ in the proof of Theorem~\ref{theo-SpecLocd0j2} are replaced by $\SF$ and $\Ind$, one obtains a proof of Theorem~\ref{tho-LSS} that is an improvement on the arguments of \cite{LSS,SS} which either needed $D_0$ to be normal or the Fredholm module to be Lipshitz regular. The key new elements in the present proof are the identity $F_\rho(D)_\rho=D_\rho$ and that $\diag(-H, FHF)$ is not restricted to $(\Hh\oplus\Hh)_\rho$, but rather deformed to $\kappa\rho D|D|^{-1}$. 
\hfill $\diamond$

\subsection{Skew localizer for $(j,d)\in\{(0,6),(4,2)\}$}
\label{sec-j0d6}

As in Section~\ref{sec-j2d0}, the index pairing is of the form $T=PFP+\one-P$, but the symmetry operator $\Ss=\Sigma S$ is odd and $\one-2P$ and $F$ are odd symmetric w.r.t. $\Ss$. The even symmetry operator $Q$ is chosen as in \eqref{eq-Q-j3d1}.

\begin{theo}
\label{theo-speclocd6j0}
For $(j,d)\in\{(0,6),(4,2)\}$, the identity \eqref{eq-SkewLocPfaf} holds with $\widehat{L}_\kappa$ given in \eqref{eq-SkewLocDef}.
\end{theo}

\noindent {\bf Proof.} Let us focus on the necessary modifications of the proof of Theorem~\ref{theo-SpecLocd0j2}. Here the $\ZM_2$-index $\Ind_2=\Ind_2(PFP+\one+P)$ is computed using Corollary \ref{coro-indes=halfflowd2j4}:
$$
\Ind_2
\;=\;
\HSF(P,F)
\;=\;
\SF(t \in [0,\tfrac{1}{2}]\mapsto  (1-t)H+tF^*HF)\;\mbox{\rm mod}\;2\;.
$$
If $\tfrac{1}{2}(H+F^*HF)$ is not invertible, $\tfrac{1}{2}(H+F^*HF)+\epsilon\one$ is invertible for $\epsilon>0$ sufficiently small. Replacing $H$ by $H+\epsilon\one$ does not change the r.h.s. of \eqref{eq-SkewLocPfaf} for $\epsilon$ sufficiently small. Thus by identifying $H$ with $H+\epsilon\one$ one may assume  without loss of generality that $\tfrac{1}{2}(H+F^*HF)$ is invertible. By Proposition \ref{prop-oddhlaf}
\begin{align*}
\HSF(P,F)
\;=\;
\Of_2\left(\imath 
\begin{pmatrix}
-H & 0\\
0 & H
\end{pmatrix},\imath 
\begin{pmatrix}
\tfrac{-1}{2}(H+F^*HF)& 0\\
0 & \tfrac{1}{2}(H+FHF^*) 
\end{pmatrix}
\right)\;.
\end{align*}
Therefore
\begin{align*}
\Ind_2
\;=\;
\Of_2\left(\imath (-H\otimes \Gamma), \imath \begin{pmatrix}
\one & 0\\
0 & F 
\end{pmatrix}(-\tfrac{1}{2}(H+F^*HF)\otimes \Gamma)\begin{pmatrix}
\one & 0\\
0 & F 
\end{pmatrix}^*\right)
\;.
\end{align*}
Let us consider the odd increasing differentiable function $F_\rho: \RM \to \RM$ defined by \eqref{eq-defFroh}.
Then $\|[F_\rho(D),H\oplus H]\|$ satisfies the bound \eqref{eq-boundFroheven}.
Moreover $F_\rho(D)$ anticommutes with $\Gamma$, hence is of the form \eqref{eq-FDform} and inherits the symmetry $Q^*\overline{F_\rho(D)}Q\;=\;-F_\rho(D)$ of $D$. As
$$
(-H\otimes \Gamma)-\begin{pmatrix}
\one & 0\\
0 & F 
\end{pmatrix}(-\tfrac{1}{2}(H+F^*HF)\otimes \Gamma)\begin{pmatrix}
\one & 0\\
0 & F 
\end{pmatrix}^*
$$ 
is compact, as in the proof of Theorem~\ref{theo-SpecLocd0j2}, the concatenation property and homotopy invariance of the orientation flow imply
\begin{align*}
\Ind_2
\;=\;
\Of_2\left(\imath \begin{pmatrix}
\one & 0\\
0 & F 
\end{pmatrix}(-\tfrac{1}{2}(H+F^*HF)\otimes \Gamma)\begin{pmatrix}
\one & 0\\
0 & F 
\end{pmatrix}^*, \imath (-H\otimes \Gamma+F_\rho(D))\right)\;.
\end{align*}
One directly checks that
\begin{align*}
s\in[0,\kappa\rho]
\;\mapsto\; 
&\begin{pmatrix}
\one & 0\\
0 & F 
\end{pmatrix}
\begin{pmatrix}
-\tfrac{1}{2}(H+F^*HF) & s\\
s & \tfrac{1}{2}(H+F^*HF) 
\end{pmatrix}
\begin{pmatrix}
\one & 0\\
0 & F 
\end{pmatrix}^*\\
&\;=\;
\begin{pmatrix}
-\tfrac{1}{2}(H+F^*HF) & sF^*\\
sF & \tfrac{1}{2}(H+FHF^*) 
\end{pmatrix}
\end{align*}
is a path of invertibles. Furthermore 
\begin{align*}
t(-H&\otimes \Gamma+\kappa F_\rho(D))\,+\,(1-t)\begin{pmatrix}
-\tfrac{1}{2}(H+F^*HF) & sF^*\\
sF & \tfrac{1}{2}(H+FHF^*) 
\end{pmatrix}\\
&\;=\;
\begin{pmatrix}
-H & t\kappa (D_0')^*+(1-t)sF^*\\
t\kappa D_0'+(1-t)sF & H
\end{pmatrix}\,+\,(t-1)\tfrac{1}{2}
\begin{pmatrix}
 F^*[H,F] & 0\\
 0 &  [H,F]F^*
\end{pmatrix}
\end{align*}
is Fredholm for all $(t,s)\in [0,1]\times [0,\kappa\rho]$, because $[H,F]$ is compact and the first summand is Fredholm by the argument leading to Theorem~\ref{theo-SpecLocd0j2}. This implies by the homotopy invariance of the orientation flow
\begin{align*}
\Ind_2&
\;=\;
\Of_2\left(\imath \begin{pmatrix}
-\tfrac{1}{2}(H+F^*HF) & \kappa\rho F^*\\
\kappa\rho F & \tfrac{1}{2}(H+FHF^*) 
\end{pmatrix}, \imath (-H\otimes \Gamma+\kappa F_\rho(D))\right)\;.
\end{align*}
As in the proof of Theorem~\ref{theo-SpecLocd0j2} the linear path connecting $(-H\otimes \Gamma+\kappa F_\rho(D))$ to $(-H\otimes \Gamma+\kappa F_\rho(D))_{\rho}\oplus (-H\otimes \Gamma+\kappa F_\rho(D))_{\rho^c}$ is within the invertibles for $\rho$ sufficiently large. Therefore
$$
\Of_2(\imath (-H\otimes \Gamma+\kappa F_\rho(D)),\imath ((-H\otimes \Gamma+\kappa F_\rho(D))_{\rho}\oplus (-H\otimes \Gamma+\kappa F_\rho(D))_{\rho^c}))\;=\;0\;.
$$
By the homotopy invariance of the orientation flow, $\Ind_2$ is equal to
\begin{align*}
\Of_2\left(\imath \begin{pmatrix}
-\tfrac{1}{2}(H+F^*HF) \!\!\!\!\!\! & \kappa\rho F^*\\
\kappa\rho F & \!\!\!\!\!\! \tfrac{1}{2}(H+FHF^*) 
\end{pmatrix}, \imath ((-H\otimes \Gamma+\kappa F_\rho(D))_{\rho}\oplus (-H\otimes \Gamma+\kappa F_\rho(D))_{\rho^c})\right)\;.
\end{align*}
Now the path
$$
s\in[0,1]
\;\mapsto\; 
A(s)\;=\;\begin{pmatrix}
-s\tfrac{1}{2}(H+F^*HF) & \kappa\rho F^*\\
\kappa\rho F & s\tfrac{1}{2}(H+FHF^*) 
\end{pmatrix}
$$
is within the invertibles for $\rho$ sufficiently large. As $tA(s)_{\rho^c}+(1-t)(-H\otimes \Gamma+\kappa F_\rho(D))_{\rho^c}$ is invertible for all $(s,t)\in[0,1]\times[0,1]$ and $\rho$ sufficiently large,
$$
tA(s)\;+\;(1-t)(-H\otimes \Gamma+\kappa F_\rho(D))_{\rho}\oplus (-H\otimes \Gamma+\kappa F_\rho(D))_{\rho^c}
$$
is Fredholm for all $(s,t)\in[0,1]\times[0,1]$. As in the proof of Theorem~\ref{theo-SpecLocd0j2} one deduces
\begin{align*}
\Ind_2
&\;=\;
\Of_2\left(\imath \kappa\rho D|D|^{-1}, \imath ((-H\otimes \Gamma+\kappa F_\rho(D))_{\rho}\oplus (-H\otimes \Gamma+\kappa F_\rho(D))_{ \rho^c})\right)\\
&\;=\;\Of_2\left(\imath \kappa\rho (D|D|^{-1})_\rho , \imath (-H\otimes \Gamma+\kappa F_\rho(D))_{\rho}\right)
\\
&\;=\;\sgn\big(\Pf(\widehat{L}_{\kappa,\rho})\big)
\,
\sgn\big(\Pf(\widehat{D}_{\rho})\big)\;,
\end{align*}
concluding the proof.
\hfill$\Box$

\subsection{Skew localizer for $(j,d)\in\{(1,0),(5,4)\}$}
\label{sec-jd10}

For $(j,d)\in\{(1,0),(5,4)\}$ one considers the index pairing $T=P\binom{F \;\;0 }{0 \;\; F} P+\one-P$ where $P=\frac{1}{2}\binom{\one\;\;-U}{-U^*\;\;\one}$. Moreover, $\Ss=\binom{\Sigma S \;\;0 }{0 \;\; -\Sigma S}$ is an even symmetry operator and $P$ is even Lagrangian and $\binom{F \;\;0 }{0 \;\; F}$ is even real  w.r.t. $\Ss$. Furthermore, $E=\frac{1}{2}\binom{\one\;\;F}{F^*\;\;\one}$ is even Lagrangian and $\binom{U \;\;0 }{0 \;\; U}$  is even real w.r.t. $\Ss$. One can build the even and odd spectral localizer as in \eqref{eq-EvenLocUnitaries} and \eqref{eq-OddLocUnitaries}. In both cases the symmetry $Q$ is of the form \eqref{eq-Q-j1d7} and its root $R$ is given by \eqref{eq-R-j1d7}. Hence one is in the situation of Section~\ref{sec-j1d7} if one considers the odd spectral localizer and in the situation of Section~\ref{sec-j2d0} if one considers the even spectral localizer.  Let us focus on the odd spectral localizer and explicitly construct the basis change $N$ leading to an off-diagonal skew localizer. Recall that $D_0$ is of the form
\begin{equation}
\label{eq-D0A2Uni}
D_0\;=\;\begin{pmatrix} 0 & D_1\\
D_1^* & 0 
\end{pmatrix}
\;,
\end{equation}
where $D_1$ is an invertible such that $D_1|D_1|^{-1}=F$. For 
\begin{equation}
\label{eq-basis2Unioff}
N\;=\;
\begin{pmatrix} 
\one & 0 & 0 & 0\\
0 & 0 & 0 & \one\\
0 & 0 & \one & 0\\
0 & \one & 0 & 0
\end{pmatrix}
\qquad
\mbox{and}
\qquad
R\;=\;
\begin{pmatrix} 
\Rr' & 0 & 0 & 0\\
0 & \Red{\imath}\Rr' & 0 & 0\\
0 & 0 & \imath\Rr' & 0\\
0 & 0 & 0 & \Red{-} \Rr'
\end{pmatrix}\;,
\end{equation}
where $\Rr' =(\Rr' )\Red{^T}$ is the unitary square root of $\Sigma S$, $N^*\widehat{L}_\kappa N$ and  $N^*\widehat{D} N$ are of the form 
$$
N^*\widehat{L}_\kappa N\;=\;\begin{pmatrix} 0 & -B^*\\
B & 0 
\end{pmatrix}
\qquad
\mbox{and}
\qquad
N^*\widehat{D} N\;=\;\begin{pmatrix} 0 & \Red{-C^*}\\
\Red{C} & 0 
\end{pmatrix}
\;,
$$
with
\begin{equation}
\label{eq-defBC}
B\;=\;
\begin{pmatrix} 
 (\Rr' ) ^*A^*\Rr'  & \kappa (\Rr' ) ^*D_1\Rr' \\
 \kappa (\Rr' ) ^*D^*_1\Rr'  & -(\Rr' ) ^*A\Rr' 
\end{pmatrix}
\qquad
\mbox{and}
\qquad
C\;=\;
\begin{pmatrix} 
 0 & (\Rr' ) ^*D_1\Rr' \\
  (\Rr' ) ^*D^*_1\Rr'  & 0
\end{pmatrix}\;.
\end{equation}
As $N$ and $(\widehat{D})^2$ commute,  $N(\widehat{\pi}_\rho)^*=(\widehat{\pi}_\rho)^*\widehat{\pi}_\rho N(\widehat{\pi}_\rho)^*$. Thus $N_\rho=\widehat{\pi}_\rho N(\widehat{\pi}_\rho)^*$ defines a real basis change on $(\Hh\oplus\Hh)^\wedge_\rho$. Then $N_\rho^*\widehat{L}_{\kappa,\rho} N_\rho$ and $N_\rho^*\widehat{D}_{\rho} N_\rho$ are of the form \eqref{eq-SkewLocOffdiag} where $B_\rho$  and $C_\rho$ are the restrictions of $B$ and $C$ to the range of $\chi(\diag(R^*D_1D_1^*R, R^*D_1^*D_1R)\leq \rho^2)$.

\begin{theo}
\label{theo-SpecLocd0j1}
For $(j,d)\in\{(1,0),(5,4)\}$, the identity \eqref{eq-SkewLocPfaf} holds with $\widehat{L}_\kappa$ build from the even or odd spectral localizer as in \eqref{eq-SkewLocDef}. Moreover, \eqref{eq-SkewLocDet} holds for $B_\rho$ and $C_\rho$ as above.
\end{theo}

\noindent {\bf Proof.}
The claim for the odd spectral localizer follows from Theorem~\ref{theo-SpecLocd3j5}, the one for the even spectral localizer from Theorem~\ref{theo-SpecLocd0j2}. The identity \eqref{eq-SkewLocDet} holds by construction of $B_\rho$ and $C_\rho$ as the Pfaffian of an off-diagional real skew-adjoint matrix equals, up to a sign depending on the size of the matrix, the determinant of its off-diagonal entry.
\hfill $\Box$

\subsection{Skew localizer for $(j,d)\in\{(3,2),(7,6)\}$}
\label{sec-jd32}

For $(j,d)\in\{(3,2),(7,6))\}$ one considers the index pairing $T=P\binom{F \;\;0 }{0 \;\; F} P+\one-P$ for $P=\frac{1}{2}\binom{\one\;\;-U}{-U^*\;\;\one}$. Moreover, $\Ss=\binom{0 \;\;\Sigma S }{\Sigma S \;\; 0}$ is an odd symmetry operator and $P$ is odd real and $\binom{F \;\;0 }{0 \;\; F}$ is odd symmetric  w.r.t. $\Ss$. Furthermore, $E=\frac{1}{2}\binom{\one\;\;F}{F^*\;\;\one}$ is odd real and $\binom{U \;\;0 }{0 \;\; U}$  is odd symmetric w.r.t. $\Ss$. One can build the even and odd spectral localizer as in \eqref{eq-EvenLocUnitaries} and \eqref{eq-OddLocUnitaries}. In both cases the symmetry $Q$ is of the form \eqref{eq-Q-j3d1}. Hence one is in the situation of Section~\ref{sec-j3d1} if one considers the odd spectral localizer and in the situation of Section~\ref{sec-j0d6} if one considers the even spectral localizer.  Let us focus on the odd spectral localizer. Again $D_0$ is of the form \eqref{eq-D0A2Uni}. \Red{After a suitable basis change $\Sigma S$ is of the form $\Sigma S=\imath \sigma_2$. Then the root $R$ of $Q$ is 
$$
R\;=\;\frac{1+i}{2}\begin{pmatrix} 
  \sigma_0 & 0 & 0 & \sigma_2\\
0 & \imath \sigma_0 & -\imath \sigma_2 & 0\\
0 & \imath \sigma_2 & \imath \sigma_0 & 0\\
-\sigma_2 & 0 & 0 & \sigma_0
\end{pmatrix}
\;.
$$}
For $N$ as in \eqref{eq-basis2Unioff},
$N^*\widehat{L}_\kappa N$ and  $N^*\widehat{D} N$ are of the form \eqref{eq-defBC} with
\begin{equation}
\label{eq-BC''}
B\;=\;
 (\Rr'')^*\,\begin{pmatrix} 
 A^* & -\kappa D_1\\
 \kappa D^*_1 & A
\end{pmatrix}\,\Rr''
\qquad
\mbox{and}
\qquad
C\;=\;
(\Rr'')^*\,\begin{pmatrix} 
 0 & -D_1\\
  D^*_1 & 0
\end{pmatrix}\,\Rr''
\end{equation}
where $\Rr''=(\Rr'')\Red{^T=\frac{1+i}{2}\binom{\;\sigma_0\;\;\sigma_2}{-\sigma_2\;\;\sigma_0}}$ is the unitary square root of $\binom{\;\;0\;\;\;\;\Sigma S}{-\Sigma S\;\;0}$. As in the previous section, $N_\rho=\widehat{\pi}_\rho N(\widehat{\pi}_\rho)^*$ defines a real basis change on $(\Hh\oplus\Hh)^\wedge_\rho$. Then $N_\rho^*\widehat{L}_{\kappa,\rho} N_\rho$ and $N_\rho^*\widehat{D}_{\rho} N_\rho$ are of the form \eqref{eq-SkewLocOffdiag} where $B_\rho$ is the restriction of $B$ to the range of $\chi((\Rr'')^*\diag(D_1D_1^*, D_1^*D_1)\Rr''\leq \rho^2)$ and $C_\rho$ is the restriction of $C$ to the same Hilbert space.

\begin{theo}
\label{theo-d2j3}
For $(j,d)\in\{(3,2),(7,6)\}$, the identity \eqref{eq-SkewLocPfaf} holds with $\widehat{L}_\kappa$ build from the even or odd spectral localizer as in \eqref{eq-SkewLocDef}. Moreover, \eqref{eq-SkewLocDet} holds for $B_\rho$ and $C_\rho$ deduced from \eqref{eq-BC''}.
\end{theo}

\noindent {\bf Proof.}
The claim for the odd spectral localizer follows from Theorem~\ref{theo-speclocd1j3}, the one for the even spectral localizer from Theorem~\ref{theo-speclocd6j0}. The identity \eqref{eq-SkewLocDet} holdy by construction of $B_\rho$ and $C_\rho$.
\hfill $\Box$

\subsection{Skew localizer for $(j,d)\in\{(0,7),(4,3)\}$}
\label{sec-jd07}

For $(j,d)\in\{(0,7),(4,3)\}$, the index pairing is $T=E(\one-2P)E+\one-E$ with $P=\chi(H\leq 0)$ being even real and $E=\chi(D_0\geq 0)$ being even Lagrangian w.r.t. the even symmetry operator $\Ss=\Sigma S$. Hence $\Ss^*\overline{H}\Ss=H$ and $\Ss^*\overline{D_0}\Ss=-D_0$. Let us use the even spectral localizer. The operator $Q$ and its root $\Rr$ satisfying \eqref{eq-ImaginaryLoc} and \eqref{eq-UnitRoot} are 
\begin{equation}
\label{eq-QRj0d7}
Q
\;=\;
\begin{pmatrix}
0 & \Ss \\ \Ss & 0
\end{pmatrix}
\;,
\qquad
R
\;=\;
\frac{1}{2}
\begin{pmatrix}
(1-\imath)\Rr & (1+\imath)\Rr \\ (1+\imath)\Rr & (1-\imath)\Rr
\end{pmatrix}
\;,
\end{equation}
where again $\Rr^2=\Ss$ and $\overline{\Rr}=\Rr^*$. The skew localizer thus is
\begin{equation}
\label{eq-SkewLoc-d3j4}
\widehat{L}_\kappa
\;=\;
\begin{pmatrix}
0 & \Rr^*(\imath\kappa D_0+ H)\Rr \\
\Rr^*(\imath\kappa D_0- H)\Rr & 0
\end{pmatrix}
\;.
\end{equation}

\begin{theo}
For $(j,d)\in\{(0,7),(4,3)\}$,
$$
\Ind_2(E(\one-2P)E+\one-E)\;=\;\sgn\big(\det(\imath\kappa D_{0,\rho}- H_\rho)\big)\,\sgn\big(\det(\imath D_{0,\rho})\big)
\;.
$$
\end{theo}

\noindent {\bf Proof.}
First let us note that the claim follows directly from \eqref{eq-SkewLocPfaf} because the skew localizer in \eqref{eq-SkewLoc-d3j4} is off-diagonal. The proof of \eqref{eq-SkewLocPfaf} can be reduced to Theorem~\ref{theo-SpecLocd3j5} in Section~\ref{sec-j1d7}. For $(j,d)\in\{(0,7),(4,3)\}$ the same symmetry relations as for $(j,d)\in\{(1,7),(5,3)\}$ hold where $\Red{\one-2P}$ is identified with $U$, and $H$ identified with $A$ (which hence has the supplementary property of being self-adjoint). By Theorem~\ref{theo-SpecLocd3j5}
$$
\Ind(E(\one-2P)E+\one-E)\;=\;
\sgn\big(\Pf((\imath R^*L^\odd_\kappa R)_\rho)\big)\,\sgn\big(\Pf((\imath R^*D^\odd R)_\rho)\big)
\;,
$$
for $\imath R^*L^{\Red{\odd}}_\kappa R$ as in \Red{\eqref{eq-SkewLoc-d3j5}}.  Because the real basis change $M$ as in \eqref{eq-PassageOddEven} commutes with $(R^*D R)^2$, $M(\widehat{\pi}_\rho)^*=(\widehat{\pi}_\rho)^*\widehat{\pi}_\rho M(\widehat{\pi}_\rho)^*$. Thus $M_\rho=\widehat{\pi}_\rho M(\widehat{\pi}_\rho)^*$ defines a real basis change on $(\Hh\oplus\Hh)^\wedge_\rho$. Then
$$
\Ind(E(\one-2P)E+\one-E)\;=\;
\sgn\big(\Pf(\imath M_\rho^*(R^*L^\odd_\kappa R)_\rho M_\rho)\big)\,\sgn\big(\Pf(\imath M_\rho^*(R^*D^\odd R)_\rho M_\rho)\big)
\;,
$$
where $\imath M_\rho^*(R^*L^\odd_\kappa R)_\rho M_\rho=-\widehat{L}_{\kappa,\rho}$ for $\widehat{L}_\kappa$ as in \eqref{eq-SkewLoc-d3j4}. As both matrices $\imath M_\rho^*(R^*L^\odd_\kappa R)_\rho M_\rho$ and $\imath M_\rho^*(R^*D^\odd R)_\rho M_\rho$ are off-diagonal, the map $\Rr_\rho:\Ran( R^*\chi(|D_0|\leq\rho) R)\to \Hh_\rho$ given by the restriction of $\Rr$ to $\Ran(R^*\chi(|D_0|\leq\rho) R)$ is unitary. Now $\Rr_\rho$ and $\Rr_\rho^*$ appear twice and the additional signs cancel out, implying the claim.
\hfill$\Box$

\subsection{Skew localizer for $(j,d)\in\{(2,1),(6,5)\}$}
\label{sec-jd21}

For $(j,d)\in\{(2,1),(6,5)\}$,the index pairing is $T=E(\one-2P)E+\one-E$ with $P=\chi(H\leq0)$ being even Lagrangian and $E=\chi(D_0\geq 0)$ being even real w.r.t. the even symmetry $\Ss=\Sigma S$. Hence $\Ss^* \overline{H}\Ss =-H$ and $\Ss^* \overline{D_0}\Ss =D_0$. Let us use the odd spectral localizer. The operator $Q$ and its root $\Rr$ satisfying \eqref{eq-ImaginaryLoc} and \eqref{eq-UnitRoot} are given as in \eqref{eq-QRj0d7}. The skew localizer thus is
\begin{equation}
\label{eq-SkewLoc-d1j2}
\widehat{L}_\kappa
\;=\;
\begin{pmatrix}
0 & \Rr^*(-\kappa D_0+\imath H)\Rr \\
\Rr^*(\kappa D_0+\imath H)\Rr & 0
\end{pmatrix}
\;.
\end{equation}

\begin{theo}
For $(j,d)\in\{(2,1),(6,5)\}$,
$$
\Ind(E(\one-2P)E+\one-E)\;=\;\sgn\big(\det(\kappa D_{0,\rho}+\imath H_\rho)\big)\,\sgn\big(\det(D_{0,\rho})\big)
\;.
$$
\end{theo}

\vspace{.2cm}

\noindent {\bf Proof.}
The claim follows directly from \eqref{eq-SkewLocPfaf} because the skew localizer in \eqref{eq-SkewLoc-d1j2} is off-diagonal. The proof of \eqref{eq-SkewLocPfaf} can be reduced to Theorem~\ref{theo-SpecLocd0j2} in Section~\ref{sec-j2d0}. For $(j,d)\in\{(2,1),(6,5)\}$ the same symmetry relations as for $(j,d)\in\{(2,0),(6,4)\}$ hold where $\one-2E$ is identified with $F$ and $D_0$ is self-adjoint. By Theorem~\ref{theo-SpecLocd0j2}
$$
\Ind(E(\one-2P)E+\one-E)\;=\;
\sgn\big(\Pf((\imath R^*L^\even_\kappa R)_\rho)\big)\,\sgn\big(\Pf((\imath R^*D^\even R)_\rho)\big)
\;,
$$
for $\imath R^*L^\even_\kappa R$ as in \eqref{eq-SkewLoc-d2j0}. Applying the real basis change $M_\rho$ as in the previous section
leads to 
$$
\Ind(E(\one-2P)E+\one-E)\;=\;
\sgn\big(\Pf(\imath M_\rho^*(R^*L^\even_\kappa R)_\rho M_\rho)\big)\,\sgn\big(\Pf(\imath M_\rho^*(R^*D^\even R)_\rho M_\rho)\big)
\;,
$$
where $\imath M_\rho^*(R^*L^\even_\kappa R)_\rho M_\rho=\widehat{L}_{\kappa,\rho}$ for $\widehat{L}_\kappa$ as in \eqref{eq-SkewLoc-d1j2}. As both matrices $\imath M_\rho^*(R^*L^\even_\kappa R)_\rho M_\rho$ and $\imath M_\rho^*(R^*D R)_\rho M_\rho$ are off-diagonal, $\Rr_\rho$   as above and $\Rr_\rho^*$ appear twice and the additional signs cancel out, this implies the claim.
\hfill$\Box$


\end{document}